\documentclass[12pt]{article}

\newcommand{\mk}[1]{\mbox{[1 mark]}}

\newcommand{\calA}{\mbox{${\cal A}$}}

\newcommand{\calD}{\mbox{${\cal D}$}}

\newcommand{\calI}{\mbox{${\cal I}$}}
\newcommand{\calM}{\mbox{${\cal M}$}}
\newcommand{\calN}{\mbox{${\cal N}$}}

\newcommand{\calR}{\mbox{${\cal R}$}}
\newcommand{\calS}{\mbox{${\cal S}$}}
\newcommand{\calT}{\mbox{${\cal T}$}}

\newcommand{\scalA}{\mbox{\scriptsize ${\cal A}$}}
\newcommand{\scalI}{\mbox{\scriptsize ${\cal I}$}}
\newcommand{\scalD}{\mbox{\scriptsize ${\cal D}$}}
\newcommand{\scalM}{\mbox{\scriptsize ${\cal M}$}}

\newcommand{\scalT}{\mbox{\scriptsize ${\cal T}$}}

\newcommand{\stLM}{\mbox{\tiny  LM}}
\newcommand{\stVLM}{\mbox{\tiny VLM}}

\def\REAL{\hbox{\rm I \kern-5.5pt R}} % my alternative
\def\NAT{\hbox{\rm I \kern-5.5pt N}} % my alternative
\def\SREAL{\hbox{\rm I \kern-3.5pt R}} % my alternative
\def\SNAT{\hbox{\rm I \kern-3.5pt N}} % my alternative

\newcommand{\logit}{\mbox{\rm logit}}

\newcommand{\bone}{{\bf 1}}

\newcommand{\bix}{\mbox{$\bm{x}$}}

\newcommand{\bB}{\mbox{\rm \bf B}}

\newcommand{\bH}{\mbox{\rm \bf H}}
\newcommand{\bI}{\mbox{\rm \bf I}}

\newcommand{\bX}{\mbox{\rm \bf X}}
\newcommand{\bW}{\mbox{\rm \bf W}}

\newcommand{\bbeta}{\mbox{\boldmath $\beta$}}

\newcommand{\boldeta}{\mbox{\boldmath $\eta$}}

\newcommand{\bomega}{\mbox{\boldmath $\omega$}}

\newcommand{\bphi}{\mbox{\boldmath $\phi$}}

\newcommand{\bpsi}{\mbox{\boldmath $\psi$}}

\newcommand{\btheta}{\mbox{\boldmath $\theta$}}

\newcommand{\Var}{ \mbox{\rm Var} }

\newcommand{\pkg}[1]{\mbox{{\sf #1}}}

\newcommand{\RR}{{\textsf{R}}}
\def\Pr{\mbox{\rm Pr}}

\newcommand{\rformula}[1]{{#1}}

\newcommand{\rcodett}[1]{{\texttt{#1}}}

\newcommand{\EIM}{\mbox{{\boldmath ${\cal I}$}${}_{\mathrm{E}}$}}   % bold \cI or \bcalI

\newcommand{\EIMsub}[1]{\mbox{{\boldmath ${\cal I}$}${}_{\mathrm{E},#1}^{}$}}

\usepackage{bm}
\usepackage{amsfonts}
\usepackage{graphicx}
\usepackage{natbib}
\usepackage{comment}

\begin{document}

\def\spacingset#1{\renewcommand{\baselinestretch}
{#1}\small\normalsize} \spacingset{1}

\title{\bf Generally--Altered, --Inflated,
  --Truncated and --Deflated Regression,
  With Application to Heaped and Seeped Data}
\author{Thomas~W.~Yee, University of Auckland\\
Chenchen Ma, Peking University
}
  \maketitle

\spacingset{1.0}

\noindent
\textbf{Abstract} \qquad
Models such as the zero-inflated and zero-altered Poisson and
zero-truncated binomial are well-established in modern
regression analysis.  We propose a super model that jointly
and maximally unifies alteration, inflation, truncation and
deflation for counts, given a 1- or 2-parameter parent (base)
distribution.  Seven disjoint sets of special value types are
accommodated because all but truncation have parametric and
nonparametric variants.  Some highlights include: (i)~the
mixture distribution is exceeding flexible, e.g., up to seven
modes; (ii)~under-, equi- and over-dispersion can be handled
using a negative binomial (NB) parent, with underdispersion
handled by a novel Generally-Truncated-Expansion method;
(iii)~overdispersion can be studied holistically in terms of
the four operators; (iv)~an important application: heaped and
seeped data from retrospective self-reported surveys are
readily handled, e.g., spikes and dips which are located
virtually anywhere; (v)~while generally-altered regression
explains why observations are there, generally-inflated
regression accounts for why they are there \textit{in excess},
and generally-deflated regression explains why observations
are \textit{not} there; (vi)~the \pkg{VGAM} \RR{} package
implements the methodology based on Fisher scoring and
multinomial logit model (Poisson, NB, zeta and logarithmic
parents are implemented.)  The GAITD-NB has potential to
become the Swiss army knife of count distributions.

\bigskip


\bigskip

\bigskip

% ---------------------------------------------------------
% \newpage  %,,,,,,,,,,,,,,,,,,,,,,,,,,,,,,,,,,,,,,,,,,,,,,,,
\section{Introduction}
\label{sec:GAITD:intro}

The analysis of counts plays an important subtopic in
regression theory.
Here, the subject of
zero-inflation,
zero-deflation,
zero-truncation,
and zero-alteration
have gained enormous traction and are now a part of
the modern regression analysis toolkit,
e.g.,
\cite{klei:zeil:2008},
\cite{zuur:save:ieno:2012},
\cite{came:triv:2013},
\cite{agre:2015},
\cite{berg:tutz:2021},
and the recent review \cite{hasl:parn:hind:mora:2022}.
In particular,
all four types of operators (``A'', ``I'', ``D'' and ``T'')
have found rich applications
in both Poisson and binomial distribution forms,
where the ZIP has been attributed to \cite{lamb:1992}
and the ZAP is often described as a hurdle model \citep{mull:1986}.
In capture--recapture experiments
the absence of~0s leads to conditional models
\citep[e.g.,][]{otis:etal:1978}
such as the positive Bernoulli distribution
or zero-truncated binomial (ZTB);
occupancy models
\citep[e.g.,][]{mack:etal:2002}
also make use of them.

Let \calR{} be the support of the
parent (base) distribution,
e.g., $\{0,1,\ldots\}$ for the Poisson.
The purpose of this paper is to extend previous work such as
the above in three directions:
\begin{itemize}\itemsep0pt

\item[(I)]
  Any subset of the support can be altered, inflated,
  deflated or truncated,
cf.~treating only the singleton~\{0\} as special.
The first three are denoted~\calA{} \calI{}, \calD{}
with finite cardinality.
The truncation set~\calT{} may be innumerable so
it is merely a proper subset of~\calR.

\item[(II)]
  Rather than allowing only one of \calA{}, \calI{},
  \calD{} and~\calT{},
the four operators are combined into a single model
and are allowed to operate concurrently. This
confers greater versatility and a holistic approach.
The \calA{}, \calI{}, \calD{} and~\calT{} are mutually disjoint.

\item[(III)]
Utilizing~(I)--(II) on~\calA{}, \calI{} and~\calD{},
parametric (subscript~``$p$'')
and nonparametric (``$np$'')
forms are spawned,
hence there are~7 special value types.
These are further combined into a `super' model,
which is informally called the GAITD `\textit{combo}'
instead for modesty.

\item[(IV)]
Although we present (I)--(III) mainly for 1- and 2-parameter 
count parents
(Poisson, negative binomial, logarithmic and zeta)
our work is envisaged for continuous distributions.

\end{itemize}
Altogether,
these directions allow a grand unification of the four
operators by the combo model
which necessitates novel methodology
such as a finite mixture distribution with \textit{nested} support.

% ,,,,,,,,,,,,,,,,,,,,,,,,,,,,,,,,,,,,,,,,,,,,,,,,
%\newpage   % ,,,,,,,,,,,,,,,,,,,,,,,,,,,,,,,,,,,,,,,,,,,,,,,,
\subsection{Some Justification}
\label{sec:GAITD:intro.motivation}

\begin{figure}[tt]
\begin{center}
 {\includegraphics[width=0.85\textwidth]{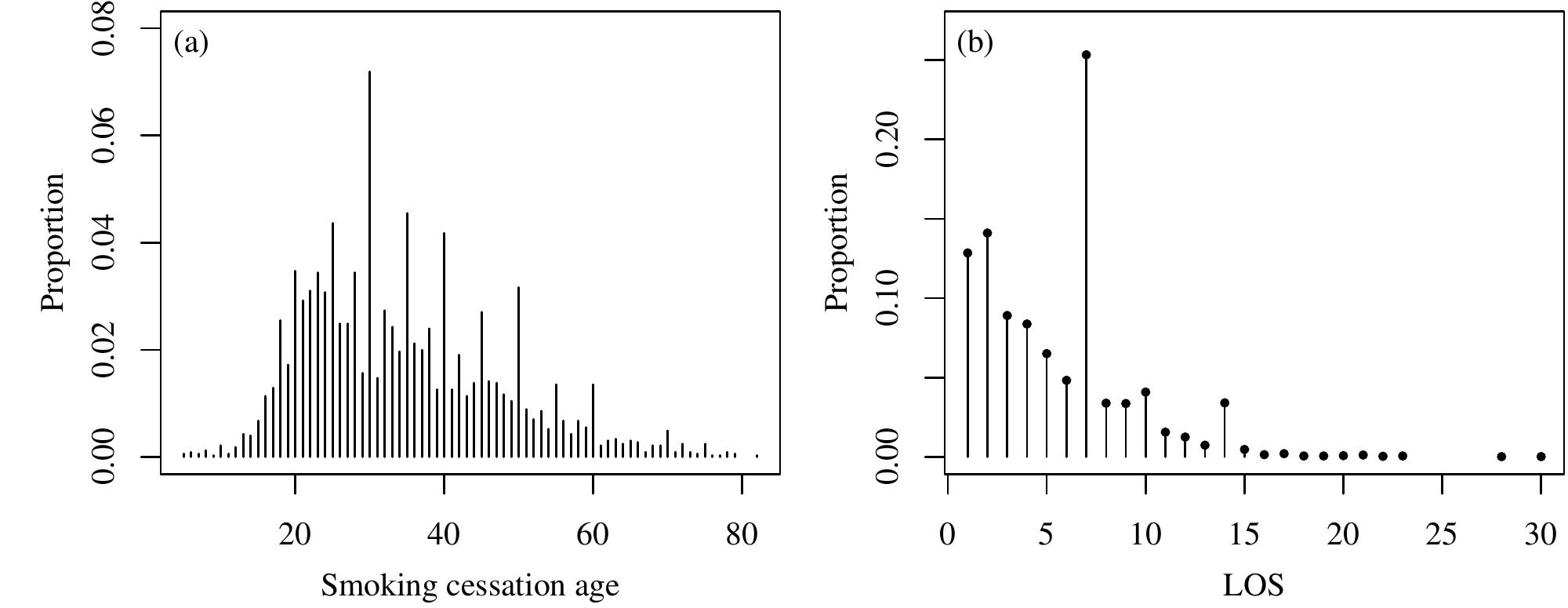}}
\end{center}
\caption{
  Two spikeplot examples from \pkg{VGAMdata}.
  (a)~Heaped variable \texttt{smokeagequit}
  in a large cross-sectional study called \texttt{xs.nz}.
(b)~Flamingo Hotel length of stay (LOS; in days, $n=4871$)
from \texttt{flamingo}.
}
 \label{fig:xsnz.heaptwoeg}
\end{figure}

\begin{figure}[tt]
\begin{center}
{\includegraphics[width=0.65\textwidth]{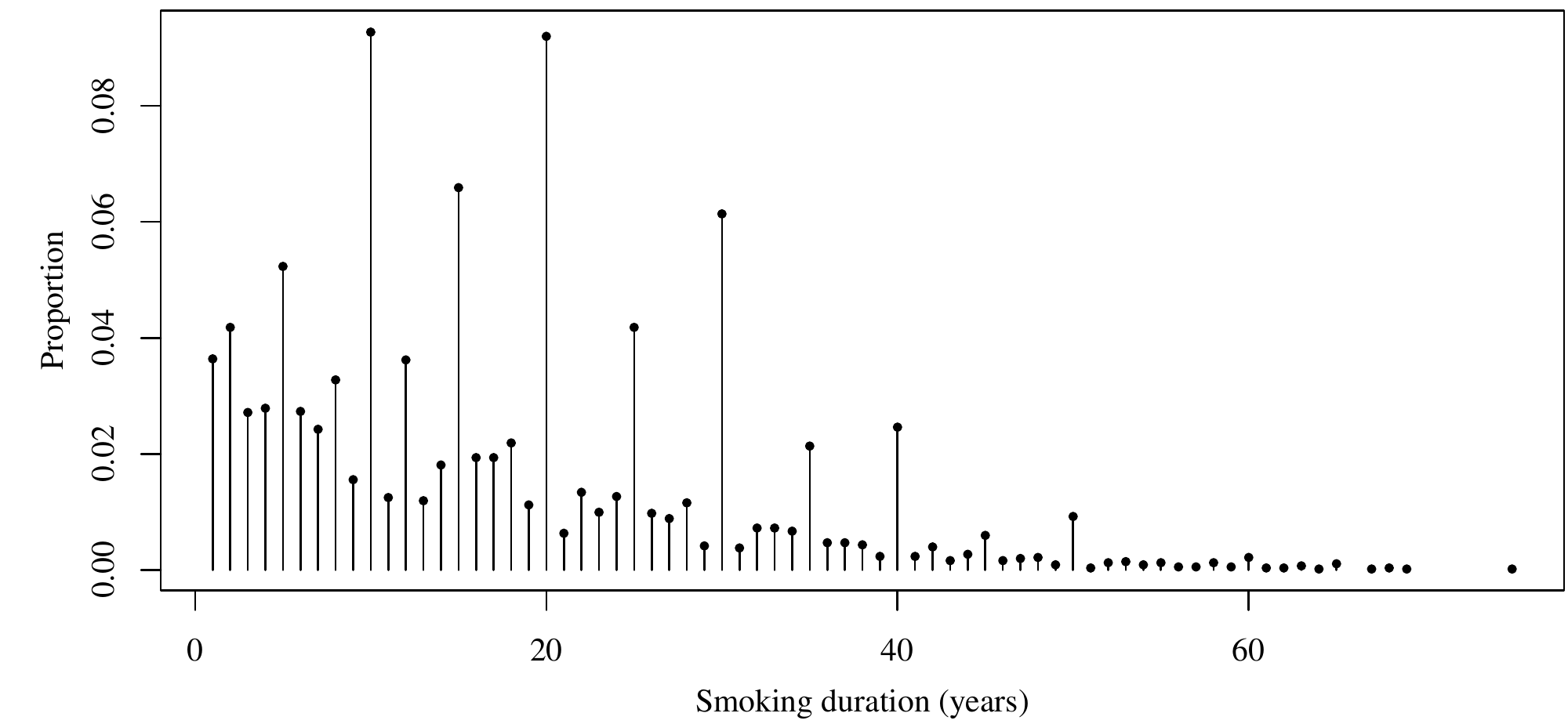}}
\end{center}
\caption{
  Smoking duration from
5492
current or past smokers in a large cross-sectional study
(Variable \texttt{smokeyears} in \texttt{xs.nz}).
The fitted GAITD regression is overlaid in
Fig.~\ref{fig:spikeplot4-smokedf}.
}
 \label{fig:xsnz.smokedur.df}
\end{figure}

Why are such extensions are so necessary?
The following short examples
illustrate why it is crucial to be able
to inflate, deflate and truncate
any set of values and not just~$\{0\}$.
\begin{itemize}\itemsep0pt

\item
 Fig.~\ref{fig:xsnz.heaptwoeg}(a)
  is a \textit{spikeplot} \citep{coxnj:2004}
  showing the proportions of self-reported age
  at which~3263 ex-smokers quit their habit.
  Two `layers' are apparent.
  The outer one is on a subset of mainly multiples of~5 and~10.
  The inner layer might be thought as being the
  `main' distribution with the outer distribution being similar
  but sampled at a greater intensity and at selected points.
  Alternatively, the  outer layer might be explained by sampling
  from the inner distribution at selected points and then added
  on top of the inner layer.

\item 
 Fig.~\ref{fig:xsnz.heaptwoeg}(b)
spikeplots the length of stay proportions of a large
4-star complex and resort located in the southern Sardinia, Italy.
Days~7 and~14 are inflated partly because of discount rates
offered to guests who book accommodation in an integral unit of weeks.
Comparing days~1 and~2, either the first day is deflated relative
to day~2, else the second day is inflated relative to day~1.

\item
With similarities to Fig.~\ref{fig:xsnz.heaptwoeg}(a),
Fig.~\ref{fig:xsnz.smokedur.df}
is a {spikeplot}
of self-reported smoking duration.
The proportions from these~$n=5492$
current or ex-smokers
appear to have a heavy-tailed distribution
such as the zeta or logarithmic but with
many multiples of~5 and~10 years having spikes.
The value~12 for a ``dozen'' may be heaped too.
A close examination shows that some inflated values
are sandwiched between two deflated values,
e.g., 29 and~31 about~30 years.
We analyze this data set in
Section~\ref{sec:GAITD:eg.xs.nz.smokeyears}.

\end{itemize}

Likewise,
applications that utilize~(I)
with respect to truncation of any set of values
is important for at least two reasons:
\begin{enumerate}\itemsep0pt

\item[(i)]
it is common to truncate the lower and/or upper tail of a
distribution.
For example, due to physical limits,
\calT{} must theoretically include~$\{25, 26, \ldots \}$
in Fig.~\ref{fig:spikeplot-xsnz-sleep}(a)
as there are only~24 hours per day.
Another compelling example is \textit{outlier deletion}:
if observations are removed then the analysis ought to
reflect this by generally-truncating those values of the support.
The Section~\ref{sec:GAITD:eg.xs.nz.sleep} analysis
is such an example;

\item[(ii)]
truncation can arise from many diverse situations.
For example, tetraphobia in East Asian culture
and triskaidekaphobia in Western culture
create structural absences in certain sampling units:
buildings that omit the 4th floor and
public passenger seating that omit row~13 are everyday examples.
Fig.~\ref{fig:xsnz.smokedur.df}
are both 0-truncated.

\end{enumerate}

As seen by two of the examples,
one very notable application of our technique is
the analysis of heaped data, an aberration ubiquitous
in surveys especially among self-reported variables
(also called `digit preference' data).
Frequently seen by an excess of
multiples of~5 and~10 relative to other values,
it is uncommon for respondents know their exact values,
hence regression analyses may suffer from bias
due to this form of measurement error
\citep{heit:rubi:1990,carr:rupp:stef:crai:2006}.
While GAITD regression can handle heaped data,
its scope is far wider since digit preference is
not the only mechanism for generating spikes,
e.g,
Fig.~\ref{fig:xsnz.heaptwoeg}(b).
Hence one source of inflation is heaped data and one source
of deflation is \textit{seeped} data---the tendency not
to select those values at the expense of the heaped values
due to measurement error.
We return to the over-/under-representation
problem of heaped/seeped data in
Section~\ref{sec:GAITD:heaped.heaping}.

% ,,,,,,,,,,,,,,,,,,,,,,,,,,,,,,,,,,,,,,,,,,,,,,,,
%\newpage   % ,,,,,,,,,,,,,,,,,,,,,,,,,,,,,,,,,,,,,,,,,,,,,,,,
\subsection{Nomenclature and notation}
\label{sec:GAITD:intro.notation}

The methodology involves four operators and draws from several
areas, therefore it is helpful to summarize most of the notation
and nomenclature used throughout here.

Nomenclaturewise due to~(II)--(III),
the acronym GAITD is used to describe the new models,
and abbreviations such as GIT for submodels
when~$\calA_{p}=\calA_{np}=\calD_{p}=\calD_{np}=\{\}$.
Any altered, inflated, deflated or truncated value is called
\textit{special} and we let~$\calS{}$ be their union.
The other values of the support are described as ordinary or
nonspecial.
Let~$\calT=\{t_1,t_2,\ldots\}$,
$\calA_{p}=\{a_1,\ldots,a_{|\scalA_{p}|}\}$,
$\calA_{np}=\{a_1,\ldots,a_{|\scalA_{np}|}\}$,
$\calI_{p}=\{i_1,\ldots,i_{|\scalI_{p}|}\}$,
$\calI_{np}=\{i_1,\ldots,i_{|\scalI_{np}|}\}$,
$\calD_{p}=\{d_1,\ldots,d_{|\scalD_{p}|}\}$ and
$\calD_{np}=\{d_1,\ldots,d_{|\scalD_{np}|}\}$
be an enumeration of the mutually exclusive
sets comprising~\calS{}.
(The notation is imperfect because~$i_1$ could
belong to~$\calI_{p}$ or~$\calI_{np}$, however the context
always renders any distinction unnecessary.)
We only allow~$|\calT|=\infty$ for upper tail truncation of the
parent; the other sets have finite cardinality.

The approach to be taken is to use modified finite mixture
distributions
\citep[e.g.,][]{fruh:cele:robe:2019}.
By `modified', the usual situation where the support
of the component distributions are all~\calR{} no longer holds.
Instead we allow them to have differing
support and sometimes they are nested and sometimes they form a
partition of~$\calR \, \backslash \, \calT$.
Although~(II) implies a single model such as a GAITD--Poisson,
as indicated in~(III)
we shall propose \textit{two} variants
which can be called, e.g.,
GAITD--Pois--MLM--MLM--MLM
and
GAITD--Pois--Pois--Pois--Pois.
Here, `MLM' stands for the \textit{multinomial logit model},
a natural extension of logistic regression to more than two classes.
The MLM variant is nonparametric because it
allows the altered, inflated or deflated
probabilities to be unstructured or unpatterned---effectively the
altered values are removed from the data set because the MLM
loosely couples with the remaining data.
The parametric variant is abbreviated
GAITD--$f_{\pi}$--$f_{\alpha}$--$f_{\iota}$--$f_{\delta}$
where~$f_{\pi}$, $f_{\alpha}$, $f_{\iota}$ and~$f_{\delta}$
are the PMFs of the
\textit{p}arent,
\textit{a}ltered,
\textit{i}nflated, and
\textit{d}eflated distributions respectively.
The parametric variant allows the altered/inflated/deflated
values to provide more information about the underlying
distribution, and in this article
these are taken to be the parent distribution itself, i.e.,
$f_{\pi} = f_{\alpha} = f_{\iota} = f_{\delta}$
but on differing support and having
potentially different parameter values.
This way, the parametric variant allows one to borrow strength across
the special values to estimate a common set of parameters,
for example.
Section~\ref{sec:GAITD:specialcases}
gives a short comparison between the two variants.
Figure~\ref{fig:Cpoz:dgaitplot.NBD.mix}(d) is
an example of a GAT-NB-MLM.

Our approach is also based on generalized linear models
\citep[GLMs;][]{neld:wedd:1972}.
The main class of models implementing GAITD regression is called
\textit{Vector Generalized Linear Models}
\citep[VGLMs;][]{yee:2015}
which are loosely multivariate GLMs
lying outside the exponential family.
They are summarized in Section~\ref{sec:Cgelimo:hdeff.VGLMs}.
Vector generalized additive models
\citep[VGAMs;][]{yee:wild:1996} and
Reduced-rank VGLMs \citep[RR-VGLMs;][]{yee:hast:2003}
offer specialized enhancements to GAITD regression analysis
but are not described here even though they may
be fitted with the same \RR{} package.

Notationally,
$\mathrm{I}(\cdot)$ denotes the indicator function, and
$\btheta = (\theta_{1},\theta_{2},\ldots,\theta_{M})^T$
for the base parameters of the parent distribution to be estimated,
and consequently~$\bbeta$ for the VGLMs regression parameters or
coefficients to be estimated.
Generically we use
$\btheta_{\pi} = (\theta_{1\pi},\theta_{2\pi},\ldots)^T$,
e.g.,
$\theta_{\pi}=\lambda_{\pi}$ for a parent Poisson,
and~$\btheta_{\alpha} = (\mu_{\alpha}, k_{\alpha})^T$
for an altered
negative binomial distribution (NBD)
having mean~$\mu_{\alpha}$ and
variance~$\mu_{\alpha} + \mu_{\alpha}^2/k_{\alpha}$.
Symbols~$\circ$ / $\otimes$ are for
the element-by-element
and Kronecker products of a matrix respectively.
The set~$\mathbb{Z}^{+}$ denotes the positive integers.

A subscript or value~$s$ is often used to index the values in
the set, e.g., $s=1,\ldots,| \calA_{np} |$
or~$s \in \calS$.
This is especially true
for parametric versus nonparametric variants
(i.e., mixture versus MLM) where we write~$\phi_p$
versus~$\phi_s,\ s=1,\ldots,|\calI_{np}|$ for instance.
Also,
let~$\omega_{\lceil a \rceil} = \{\omega_s: a_s = a\}$
so that~$\sum_{a \in \scalA_{np}} \, a \, \omega_{\lceil a \rceil}$
is equivalent to
$\sum_{s=1}^{|\scalA_{np}|} \, a_s \, \omega_{s}$---sometimes one form
is preferable.
Similarly~$\phi_{\lceil i \rceil}$
and~$\psi_{\lceil d \rceil}$ for inflated
and deflated distributions respectively.

\begin{figure}[hh]
\begin{center}
 {\includegraphics[width=0.70\textwidth]{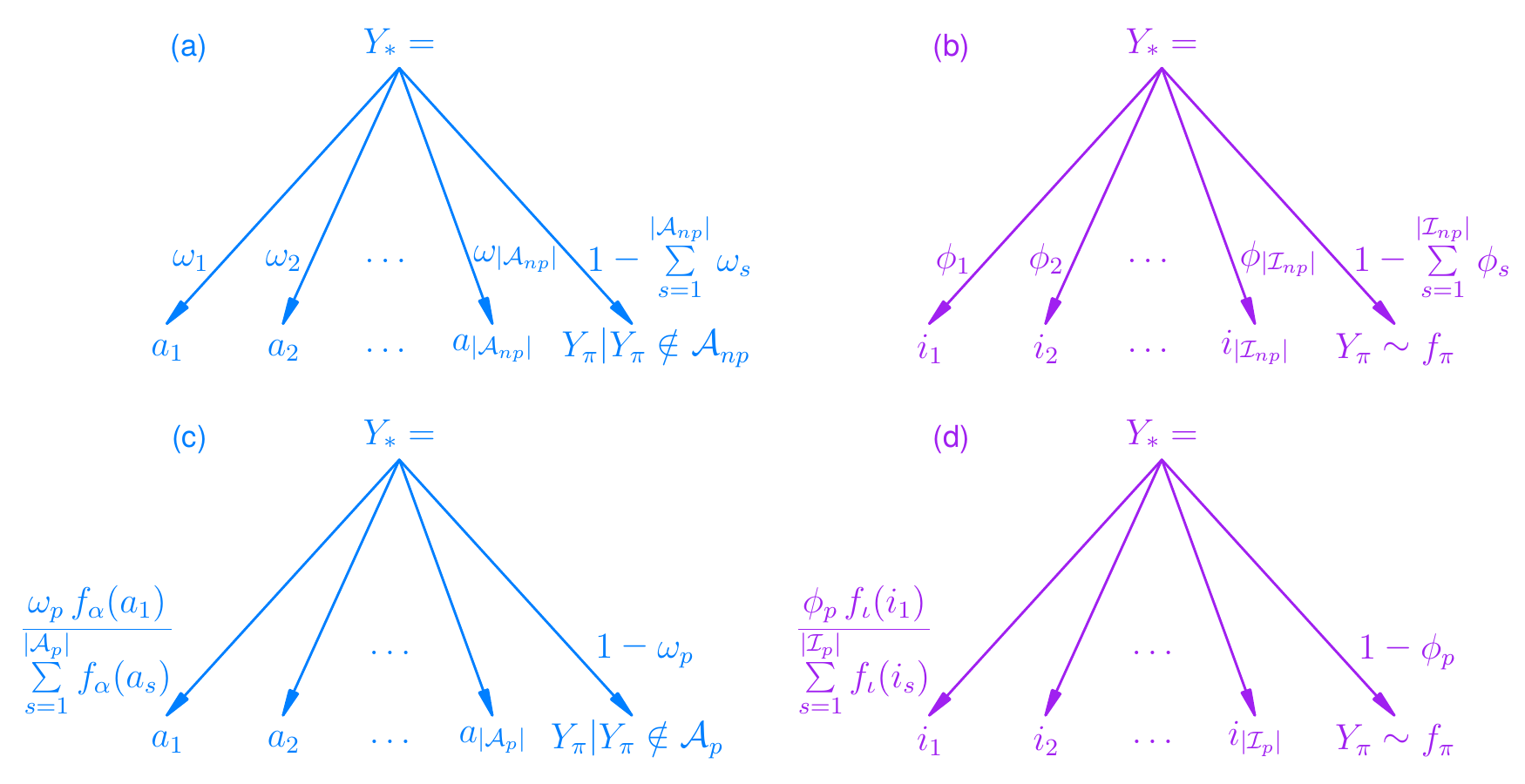}}
\end{center}
\caption{Decision tree diagram for
generally-altered
((a), (c))
and
generally-inflated
((b), (d))
distributions,
abbreviated as
(a)~GA--$f_{\pi}$--MLM,
(b)~GI--$f_{\pi}$--MLM,
(c)~GA--$f_{\pi}$--$f_{\alpha}$,
(d)~GI--$f_{\pi}$--$f_{\iota}$,
so that (a)--(b) are nonparametric and
(c)--(d) are parametric.
Here, $f_{\pi}$ is the parent distribution PMF,
and $f_{\alpha}$ and $f_{\iota}$ are PMFs for
the altered and inflated values.
Also, \rformula{$Y_{\pi}$} corresponds to the parent distribution
and~\rformula{$Y_*$} is the response of interest.
For (a)--(b) the positive
probabilities~\rformula{$\omega_s$} and~\rformula{$\phi_s$}
satisfy $0 < \sum_{s=1}^{| \scalA_{np} |} \omega_s < 1$
and~$0 < \sum_{s=1}^{| \scalI_{np} |} \phi_s < 1$
and are modelled by an MLM.
In~(c) $\Pr(Y_* = a_j) =
\omega_p \, f_{\alpha}(a_j) / {\sum_{s=1}^{| \scalA_{p} |} f_{\alpha}(a_s)}$;
in~(d) $\Pr(Y_* = i_j) =
\phi_p   \, f_{\iota}(i_j) / {\sum_{s=1}^{| \scalI_{p} |} f_{\iota}(i_s)} +
(1-\phi_p) \, f_{\pi}(i_j)$.
\label{fig:flowchart.gait.4}
}
\end{figure}

% ---------------------------------------------------------
% \clearpage  % ,,,,,,,,,,,,,,,,,,,,,,,,,,,,,,,,,,,,,,,,,,,,,,,,
% \newpage  % ,,,,,,,,,,,,,,,,,,,,,,,,,,,,,,,,,,,,,,,,,,,,,,,,
\subsection{Vector Generalized Linear Models}
\label{sec:Cgelimo:hdeff.VGLMs}

As GAITD regression is fitted as a VGLM via~(\ref{eq:gaitd.etas})
they are briefly summarize here.
Let the dimension of covariates~$\bix_i$ be~$d$ with~$x_1=1$
denoting the optional intercept.
The log-likelihood is~$\ell=\sum_{i=1}^n w_i^* \ell_i$
where the prior weights~$w_i^*$ are positive, known and prespecified.
VGLMs use multiple linear predictors~$\eta_j$
to model multiple parameters.

For $M$ parameters~$\theta_{j}$
VGLMs specify the~$j$th~linear predictor as
\begin{equation}
g_j(\theta_{j}) ~=~
\eta_j ~=~ \bbeta_j^{T} \bix ~=~
\sum_{k=1}^d \; \beta_{(j)k} \, x_{k}, ~~~ j=1, \ldots, M,
\label{gammod2}
\end{equation}
for some suitable \textrm{parameter link function}~$g_j$
satisfying the usual properties.
For example, the NBD as a VGLM has
$\eta_1=\log\mu$ and
$\eta_2=\log k$ by default.
Since~$M>1$ linear constraints between
the regression coefficients are accommodated by
\begin{eqnarray}
\boldeta(\bix_i)
  & = &
\left( \eta_1(\bix_i), \ldots, \eta_M(\bix_i) \right)^T ~=~
\sum_{k=1}^d \, \bbeta_{(k)} \, x_{ik} ~=~
\sum_{k=1}^d \, \bH_k \, \bbeta_{(k)}^{*} \, x_{ik} ~=~
\bB^T \bix_{i},  ~~
\label{eq:constraints.VGLM}
\end{eqnarray}
for known \textit{constraint matrices}~$\bH_k$
of full column-rank
(i.e., rank~$R_k=$ \rcodett{ncol}($\bH_k$)),
and~$\bbeta_{(k)}^{*}$
is a possibly reduced set of regression coefficients
to be estimated.
While trivial constraints are denoted by~$\bH_k=\bI_M$,
other common examples include parallelism ($\bH_k=\bone_M$),
exchangeability,
intercept-only parameters~$\eta_j=\beta_{(j)1}^*$, and
selecting different subsets of~$\bix_i$ for modelling each~$\eta_j$.
The overall `large' model matrix is~$\bX_{\stVLM}$, which
is~$\bX_{\stLM} \otimes \bI_M$ with trivial constraints,
while~$\bX_{\stLM}=[(x_{ik})]$ is the `smaller'~$n \times d$ model
matrix associated with a $M=1$ model.

As with GLMs,
iteratively reweighted least squares (IRLS)/Fisher scoring
is the central algorithm for VGLMs.
Consequently the score vector and expected
information matrix (\EIM; EIM) are needed
(see the Supplementary Materials).
In particular,
let~$\bW^{(a)}=(\bW_1^{(a)},\ldots,\bW_n^{(a)})$ be the working
weight matrices, comprising~$\bW_i^{(a)} = \mbox{}$
$-\mathrm{E}
[\partial^2 \ell_i / (\partial \boldeta_i\,\partial \boldeta_i^T)]$
at iteration~$a$.
Fisher scoring has the~$\bW_i$ as
$\EIMsub{i} \circ
  \left[
  ({\partial \btheta  }/{\partial \boldeta_i  })
  ({\partial \btheta^T}/{\partial \boldeta_i^T})
  \right]$.
That is,
\begin{eqnarray}
\label{eq:hdeff:eim.ab}
(\bW_i)_{uv}
&=&
-\mathrm{E}
  \left[
  \frac{ \partial^2 \ell_i}{\partial \eta_u\,\partial \eta_v}
  \right] ~=~
 \mathrm{E}
  \left[
  \frac{-\partial^2 \ell_i}{\partial \theta_u\,\partial \theta_v}
  \right]\;
  \frac{\partial \theta_u}{\partial \eta_u} \;
      \frac{\partial \theta_v}{\partial \eta_v}
  ~=~
(\EIMsub{i})_{uv} \cdot
\left[ \; g'_u(\theta_u) \; \, g'_v(\theta_v) \right]^{-1},~~~
\end{eqnarray}
say,
for~$u,v\in\{1,\ldots,M\}$.
In particular, (\ref{eq:hdeff:eim.ab}) holds for 1-parameter
link functions~$g_u$.
The estimated variance-covariance matrix is
$\widehat{\Var}\left( \widehat{\bbeta^{*}} \right) =
\left( \bX_{\stVLM}^T \; \widehat{\bW} \; \bX_{\stVLM} \right)^{-1}$
evaluated at the MLE,
where~$\bbeta^{*}=(\bbeta_{(1)}^{*T},\ldots,\bbeta_{(d)}^{*T})^T$
are all the regression coefficients to be estimated.

\begin{figure}[tt]
\begin{center}
{\includegraphics[width=1.0\textwidth]{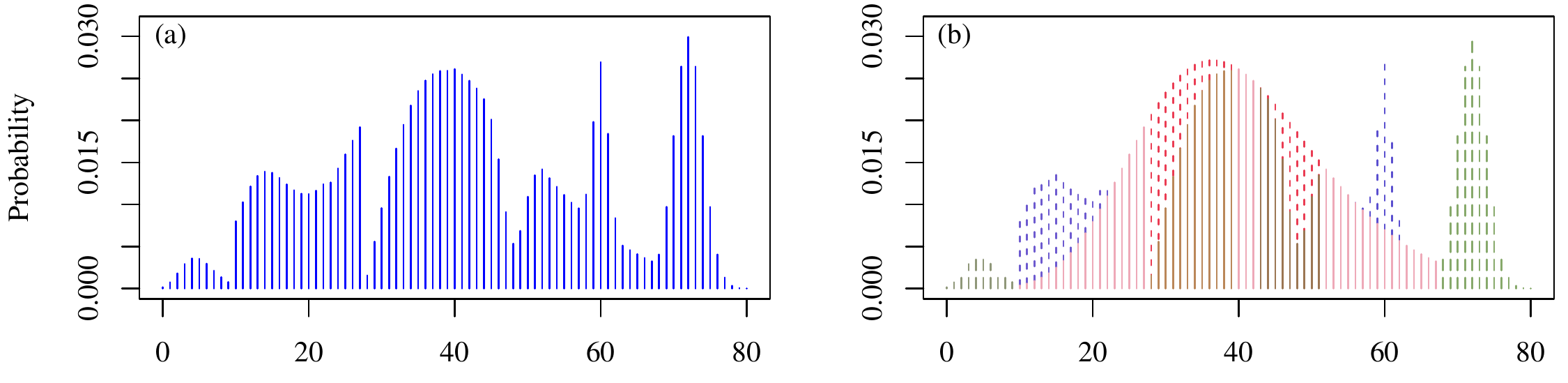}}
\end{center}
\caption{
A GAITD--NB distribution with seven modes;
(a)~overall masked PMF;
(b)~PMF decomposed by the special values
using color and various line types,
e.g.,
the dip probabilities appear in reddish dashed lines.
\label{fig:Cpoz:gaitd.nb.modes7}
}
\end{figure}

% =========================================================
%\newpage  % ,,,,,,,,,,,,,,,,,,,,,,,,,,,,,,,,,,,,,,,,,,,,,,,,
\section{The GAITD `Combo'  Model}
\label{sec:GAITD:combo}

% ---------------------------------------------------------
%\newpage  % ,,,,,,,,,,,,,,,,,,,,,,,,,,,,,,,,,,,,,,,,,,,,,,,,
\subsection{Probability Mass Function}
\label{sec:GAITDcombo.pmf}

The (parametric and nonparametric) GAITD combo PMF is

\noindent
$\Pr(Y_{*} = y; \btheta_{\pi},
\omega_{p}, \btheta_{\alpha},
\phi_{p},   \btheta_{\iota},
\psi_{p},   \btheta_{\delta},
\bomega_{np},
\bphi_{np},
\bpsi_{np})
= f(y) = \mbox{}$
\begin{eqnarray}
\left\{
\begin{array}{ll}
0, &
y \in \calT,
\\
% ,,,,,,,,,,,,,,,,,,,,,,,,,,,,,,,,,,,,,,,,,,,,,,,,,,,,,,,,,
  \omega_p \,
  {f_{\alpha}(y)} / {
  \sum\limits_{u \in \scalA_p} f_{\alpha}(u)}, ~~~~~ &
y \in \calA_{p},
%\label{eq:dgaitd.Ap}
\\
% ,,,,,,,,,,,,,,,,,,,,,,,,,,,,,,,,,,,,,,,,,,,,,,,,,,,,,,,,,
  \omega_s, ~~~~~
& y = a_s \in
 \calA_{np},\ s=1,\ldots,|{\calA_{np}}|,  %L_{A_{np}},~~~~
%\label{eq:dgaitd.Anp}
\\
% ,,,,,,,,,,,,,,,,,,,,,,,,,,,,,,,,,,,,,,,,,,,,,,,,,,,,,,,,,
\Delta \, f_{\pi}(y) + \phi_p \,
f_{\iota}(y) / {\sum\limits_{u \in \scalI_p} f_{\iota}(u)}, ~~~~~ &
y \in \calI_p,
%\label{eq:dgait.Ip}
\\
% ,,,,,,,,,,,,,,,,,,,,,,,,,,,,,,,,,,,,,,,,,,,,,,,,,,,,,,,,,
\Delta \, f_{\pi}(y) + \phi_s, ~~~~~
& y = i_s \in \calI_{np},\ s=1,\ldots,|{\calI_{np}}|,  %L_{I_{np}},
\qquad
%\label{eq:dgaitd.Inp}
\\
% ,,,,,,,,,,,,,,,,,,,,,,,,,,,,,,,,,,,,,,,,,,,,,,,,,,,,,,,,,
\Delta \, f_{\pi}(y) - \psi_p \,
f_{\delta}(y) / \sum\limits_{u \in \scalD_p} f_{\delta}(u), ~~~~~ &
y \in \calD_p,
\\
% ,,,,,,,,,,,,,,,,,,,,,,,,,,,,,,,,,,,,,,,,,,,,,,,,,,,,,,,,,
\Delta \, f_{\pi}(y) -\psi_s, ~~~~~
& y = d_s \in \calD_{np},\ s=1,\ldots,|{\calD_{np}}|,
\qquad
\\
% ,,,,,,,,,,,,,,,,,,,,,,,,,,,,,,,,,,,,,,,,,,,,,,,,,,,,,,,,,
\Delta \, f_{\pi}(y),  &
y \in \calR \backslash \calS,
% ,,,,,,,,,,,,,,,,,,,,,,,,,,,,,,,,,,,,,,,,,,,,,,,,,,,,,,,,,
\end{array}
\right.
\label{eq:dgaitdpmf.combo}
\end{eqnarray}
where the normalizing constant is
\begin{eqnarray}
  \Delta
  &~=~&
\frac{
 1 - \omega_p - \phi_p + \psi_p
   - \sum\limits_{u=1}^{| \scalA_{np} |}  \omega_u
   - \sum\limits_{u=1}^{| \scalI_{np} |}  \phi_u
   + \sum\limits_{u=1}^{| \scalD_{np} |}  \psi_u
}{
1 -
 \sum\limits_{a \in \{ \scalA_p,\; \scalA_{np}\} } f_{\pi}(a) -
 \sum\limits_{t \in \scalT} f_{\pi}(t)
} .
\label{eq:dgaitdpmf.combo.mux}
\end{eqnarray}
Alternatively, (\ref{eq:dgaitdpmf.combo}) might be called
the \textit{complete} or \textit{full} GAITD model.
It is fully specified\\
GA${}_{p}$A${}_{np}$I${}_{p}$I${}_{np}$D${}_{p}$D${}_{np}$T--\nobreak
$f_{\pi}(\calR, \btheta_{\pi})$--\nobreak
$f_{\alpha}(\calA_p, \omega_p, \btheta_{\alpha})$--\nobreak
MLM($\calA_{np}$, $\bomega_{np}$)--\nobreak
$f_{\iota}(\calI_p, \phi_p, \btheta_{\iota})$--\nobreak
MLM($\calI_{np}$, $\bphi_{np}$)--\nobreak
$f_{\delta}(\calD_p, \psi_p, \btheta_{\delta})$--\nobreak
MLM($\calD_{np}$, $\bpsi_{np}$)--\calT.
The~$\btheta_{\alpha}$ are asymptotically orthogonal to all the others.

Equation~(\ref{eq:dgaitdpmf.combo}) has a simple structure.  The
term~$\Delta \, f_{\pi}(y)$ is the scaled parent distribution from
which the inflated and deflated values have probabilities added to
or subtracted from to produce spikes or dips.  In general,
deflation can be thought of as the opposite of inflation.
On~$\calA_{np}$ the altered probabilities~$\omega_s$
are ordinary probabilities estimated from the data which
do not emanate from the scaled parent
directly---Fig.~\ref{fig:Cpoz:dgaitplot.NBD.mix}(d)
of a GAT-NB-MLM is an example.
Elements from~$\calA_p$ have a conditional~$f_{\alpha}$
distribution on that subset.

Fig.~\ref{fig:flowchart.gait.4} depicts flowcharts for
four types of GA-- and GI-- submodels where
plots~(b) and~(d) show the two-source concept for
inflated values: (b)~is nonparametric because
the~$\phi_s$ are unstructured
while~(d) is parametric because
the additional probabilities are realizations from~$f_{\iota}$
on a subset of its support.
Plots~(a) and~(c) show how altered values only have a
single source and repeat the same parametric versus
nonparametric variant idea in GI submodels.

The precedence of the operators is
truncation,
alteration,
inflation and lastly
deflation,
to avoid potential interference among them.
Each line of~(\ref{eq:dgaitdpmf.combo}) corresponds to a special
value type except for the final line for the nonspecial values.
It is clear that
alteration occurs on a partitioned support whereas
inflation and deflation occur on \textit{nested} support.
The parametric variants fit an additional distribution
compared to the parent.
The nonparametric variants entail estimating unstructured
probabilities~$\omega_s$, $\phi_s$ and~$\psi_s$
as freely as possible using a MLM.

GAITD count distributions are exceedingly flexible, for example,
they can accommodate up to seven modes,
e.g.,
Fig.~\ref{fig:Cpoz:gaitd.nb.modes7} uses a NB
parent where the first plot is the overall distribution and the
second unmasks each finite mixture by color and line type.

% ,,,,,,,,,,,,,,,,,,,,,,,,,,,,,,,,,,,,,,,,,,,,,,,,
%\newpage   % ,,,,,,,,,,,,,,,,,,,,,,,,,,,,,,,,,,,,,,,,,,,,,,,,
\subsection{Goals and Inference}
\label{sec:GAITD:inference}

Based on~(\ref{eq:dgaitdpmf.combo}),
three fundamental questions that may be answered by
GAITD regression are as follows.
For concreteness, suppose age and sex are covariates.
\begin{enumerate}\itemsep0pt

\item[(1)]
  For alteration,
  the probabilities exist in the usual form of
  an ordinary quantity~$\omega_s$ or~$\omega_p$,
  hence generally-altered regression
  explains why observations are there,
  e.g.,
  which covariates explain~$\Pr(Y=a_s)$?
  How do age and sex affect~$\Pr(Y=a_s)$?

\item[(2)]
For inflation,
the~$\phi_s$ and~$\phi_p$ are additions to the
scaled parent distribution,
hence may be called \textit{structural} probabilities
following~$\phi$ being referred to
as the probability of a structural~0 for the ZIP.
Alternatively, the~$\phi_s$ and~$\phi_p$
may be called the \textit{excess}
(from extreme value terminology)
or
mixing probabilities.
Generally-inflated regression accounts for why observations are
there \textit{in excess},
e.g., older males have a greater chance of being
represented at certain values of~$i_s$.
What other variables contribute to overrepresentation at~$i_s$?

\item[(3)]
For deflation,
the~$\psi_s$ and~$\psi_p$ may be called \textit{dip} probabilities
because they are subtracted from the scaled parent distribution.
Alternative names might be
the \textit{shortfall} or \textit{deficit} probabilities.
Generally-deflated regression explains why observations are
\textit{not} there,
e.g., younger females may be underrepresented in the
data at certain values of~$d_s$.

\end{enumerate}
Furthermore,
while nonparametric
GA--, GI-- and GD-- analysis at specific special values may 
be of interest in their own right,
these analysis types may be used to deal
with \textit{nuisance values}---aberrant values that are not
of interest but nevertheless must be adjusted for.
(This is because the nonparametric variants model unstructured
probabilities.)
Hence the additional goals are:
\begin{enumerate}\itemsep0pt

\item[(4)]
  nonparametric
  general-alteration may be used to `delete' values equalling~$a_s$
  because some uninteresting probability~$\omega_s$ is
  used to estimated it;

\item[(5)]
  nonparametric
  general-inflation may be used to shave off the spikes so that
  inference may be directed at~$f_{\pi}$;

\item[(6)]
  nonparametric
  general-deflation may be used to fill in the dips
  (cracks or nonexisting values)
  so that inference may be focussed on~$f_{\pi}$.

\end{enumerate}
In short,
nonparametric analyses disentangle various aberrations
in the data
to allow inference on the underlying parent distribution.

%==============================================================
% \newpage  % ,,,,,,,,,,,,,,,,,,,,,,,,,,,,,,,,,,,,,,,,,,,,,,,,,
\subsection{Multinomial Logit Model and Identifiability}
\label{sec:gaitd:MLM.regconds}

The MLM for a probability~$p_s$ has inverse link (softmax)
of the form~$e^{\eta_{s}} / \sum\limits_{m \in \scalM} e^{\eta_{m}}$
for some set~$\calM$
and~$\eta_{1+|\scalM|} \equiv 0$ by default.
(Further details are in the Supplementary Materials.)
For 1-parameter distributions,
the VGLM formulation~(\ref{eq:constraints.VGLM})
of~(\ref{eq:dgaitdpmf.combo}) is $\boldeta^T = \mbox{}$
\begin{eqnarray}
\nonumber
&&
\Bigg( g_{\pi}(\theta_{\pi}), ~
\log \frac{\omega_p}{\calN}, ~ g_{\alpha}(\theta_{\alpha}), ~
\log \frac{\phi_p}{\calN}, ~ g_{\iota}(\theta_{\iota}), ~
\log \frac{\psi_p}{\calN}, ~ g_{\delta}(\theta_{\delta}), ~
\log \frac{\omega_1}{\calN},\ \ldots, ~
\log \frac{\omega_{|\scalA_{np}|}}{\calN},
  ~ \qquad \\
&&  ~~~~~~
\log \frac{\phi_1}{\calN},\ \ldots, ~
\log \frac{\phi_{|\scalI_{np}|}}{\calN}, ~
\log \frac{\psi_1}{\calN},\ \ldots, ~
\log \frac{\psi_{|\scalD_{np}|}}{\calN}
\Bigg).  ~~~
  \label{eq:gaitd.etas}
\end{eqnarray}
where
$\calN =  1 - \omega_p - \phi_p - \psi_p
    - \sum\limits_{u=1}^{| \scalA_{np} |}  \omega_u
    - \sum\limits_{u=1}^{| \scalI_{np} |}  \phi_u
    - \sum\limits_{u=1}^{| \scalD_{np} |}  \psi_u$
is the baseline probability.
The ordering keeps the unstructured probabilities contiguous
which simplifies the implementation.
All but four linear predictors are coupled together by the MLM.
The VGLM estimates all the parameters and probabilities
by the regression coefficients in~(\ref{eq:constraints.VGLM}).

Disallowing degeneracy, the constraints on the parameter
space needed for~(\ref{eq:dgaitdpmf.combo})
to be identifiable are
\begin{eqnarray}
  \label{eq:dgaitdpmf.combo.constraints1}
\begin{array}{ll}
% ,,,,,,,,,,,,,,,,,,,,,,,,,,,,,,,,,,,,,
0 < \phi_s     \mathrm{~for~} s=1,\ldots,|\calI_{np}|, ~~~~~~
&
0 < \phi_p   ,
\\
% ,,,,,,,,,,,,,,,,,,,,,,,,,,,,,,,,,,,,,
0 < \psi_s     \mathrm{~for~} s=1,\ldots,|\calD_{np}|, ~~~~~~
&
0 < \psi_p   ,
\\
% ,,,,,,,,,,,,,,,,,,,,,,,,,,,,,,,,,,,,,
0 < \omega_s  \mathrm{~for~}  s=1,\ldots,|\calA_{np}|,
&
0 < \omega_p , ~~~~~~
\\[.1cm]
% ,,,,,,,,,,,,,,,,,,,,,,,,,,,,,,,,,,,,,
% ,,,,,,,,,,,,,,,,,,,,,,,,,,,,,,,,,,,,,
\phi_p - \psi_p + \omega_p
   + \sum\limits_{s=1}^{| \scalI_{np} |}  \phi_s
   - \sum\limits_{s=1}^{| \scalD_{np} |}  \psi_s
   + \sum\limits_{s=1}^{| \scalA_{np} |}  \omega_s < 1, ~~~~
&
% ,,,,,,,,,,,,,,,,,,,,,,,,,,,,,,,,,,,,,
|\calR \backslash \calS | > 0.
\end{array}
\end{eqnarray}
The last condition
ensures that the entire support cannot be
inflated, deflated, altered or truncated,
and guarantees that~$\dim(\btheta_{\pi}) \geq 1$;
and if~$\widetilde{\calR}$ is the support of the sample
(i.e., set of all response values)
then~$| \widetilde{\calR} \backslash \calS | > 0$ must hold too.
Practically however,
the number of nonspecial values should exceed unity
to avoid a trivial regression.
Note that
$\omega_p=0$ is not permitted because otherwise~$\calA_p$
could be subsumed into~\calT,
and a similar argument holds for $\omega_s=0$,
as well as for
$\phi_p=0$, $\phi_s=0$,
$\psi_p=0$ and $\psi_s=0$.

Continuing with identifiability issues,
the ZAP can arise in two ways:
either $\calA_{p}=\{0\}$ or $\calA_{np}=\{0\}$;
and likewise
$\calI_{p}=\{0\}$ or $\calI_{np}=\{0\}$ for the ZIP.
To ensure the parameters are identifiable one can further enforce
\begin{eqnarray}
  \label{eq:dgaitdpmf.combo.constraints2}
|\calA_{p}| \neq 1,  ~~~~~
|\calI_{p}| \neq 1
\mathrm{~~and~~}
|\calD_{p}| \neq 1.
\end{eqnarray}

% =========================================================
%\newpage  % ,,,,,,,,,,,,,,,,,,,,,,,,,,,,,,,,,,,,,,,,,,,,,,,,
\section{GAITD Distributions: Properties \& Applications}
\label{sec:GAITD:properties}

We present
some basic distributional properties for the combo
model and special cases.

% ,,,,,,,,,,,,,,,,,,,,,,,,,,,,,,,,,,,,,,,,,,,,,,,,
%\newpage   % ,,,,,,,,,,,,,,,,,,,,,,,,,,,,,,,,,,,,,,,,,,,,,,,,
\subsection{Moments and CDF}
\label{sec:GAITD:combo.moments.cdf}

For the combo mean and variance
the~$k$th moment is

\begin{eqnarray}
\label{eq:gait.combo.EYk}
\mathrm{E}[Y_{*}^{k}]  &=&
\frac{ \omega_p \,
\sum\limits_{a \in \scalA_p} \, a^k \, f_{\alpha}(a)
}{
\sum\limits_{a \in \scalA_p} \, f_{\alpha}(a)
}
+
\frac{ \phi_p \,
\sum\limits_{i \in \scalI_p} \, i^k \, f_{\iota}(i)
}{
\sum\limits_{i \in \scalI_p} \, f_{\iota}(i)
}
-
\frac{ \psi_p \,
\sum\limits_{d \in \scalD_p} \, d^k \, f_{\delta}(d)
}{
\sum\limits_{d \in \scalD_p} \, f_{\delta}(d)
}
+
\sum\limits_{a \in \mathcal{A}_{np}} a^k \, \omega_{\lceil a \rceil} +
\mbox{}  
\\ &&
\nonumber
\sum\limits_{i \in \mathcal{I}_{np}} i^k \,   \phi_{\lceil i \rceil} -
\sum\limits_{d \in \mathcal{D}_{np}} d^k \,   \psi_{\lceil d \rceil} +
\Delta
\cdot
\left\{
\mathrm{E}[Y_{\pi}^k]
- \sum\limits_{t \in \scalT} \, t^k \, f_{\pi}(t)
- \sum\limits_{a \in \{\scalA_p,\; \scalA_{np}\}} a^k \, f_{\pi}(a)
\right\}.
\end{eqnarray}
Let $F_*$ be the GAITD cumulative distribution function (CDF)
and $F_{\pi}$ the CDF of the parent distribution.
Then
\begin{eqnarray}
\nonumber
F_*(y) &=&
\omega_p \;
\frac{ \sum\limits_{a \in \mathcal{A}_p} \mathrm{I}(a \leq y) \,
f_{\alpha}(a)}{
\sum\limits_{a \in \scalA_p} \; f_{\alpha}(a)
}
+
\phi_p \;
\frac{\sum\limits_{i \in \mathcal{I}_p} \mathrm{I}(i \leq y) \,
f_{\iota}(i)}{
\sum\limits_{i \in \scalI_p} \, f_{\iota}(i)
}
-
\psi_p \;
\frac{\sum\limits_{d \in \mathcal{D}_p} \mathrm{I}(d \leq y) \,
f_{\delta}(d)}{
\sum\limits_{d \in \scalD_p} \, f_{\delta}(d)
}
+
\mbox{} \\ &&
\nonumber
\sum\limits_{a \in \mathcal{A}_{np}} \; \mathrm{I}(a\leq y) \,
\omega_{\lceil a \rceil} +
\sum\limits_{i \in \mathcal{I}_{np}} \; \mathrm{I}(i\leq y) \,
\phi_{\lceil i \rceil} -
\sum\limits_{d \in \mathcal{D}_{np}} \; \mathrm{I}(d\leq y) \,
\psi_{\lceil d \rceil} +
\mbox{} \\ &&
\Delta \cdot
\left\{
F_{\pi}(y)
- \sum\limits_{t \in \mathcal{T}} \; \mathrm{I}(t \leq y) \, f_{\pi}(t)
- \sum\limits_{a \in \{\mathcal{A}_p,\, \mathcal{A}_{np}\}}
  \; \mathrm{I}(a \leq y) \, f_{\pi}(a)
\right\}.
\label{eq:pgait.combo}
\end{eqnarray}

\begin{figure}[tt]
\begin{center}
 {\includegraphics[width=0.8\textwidth]{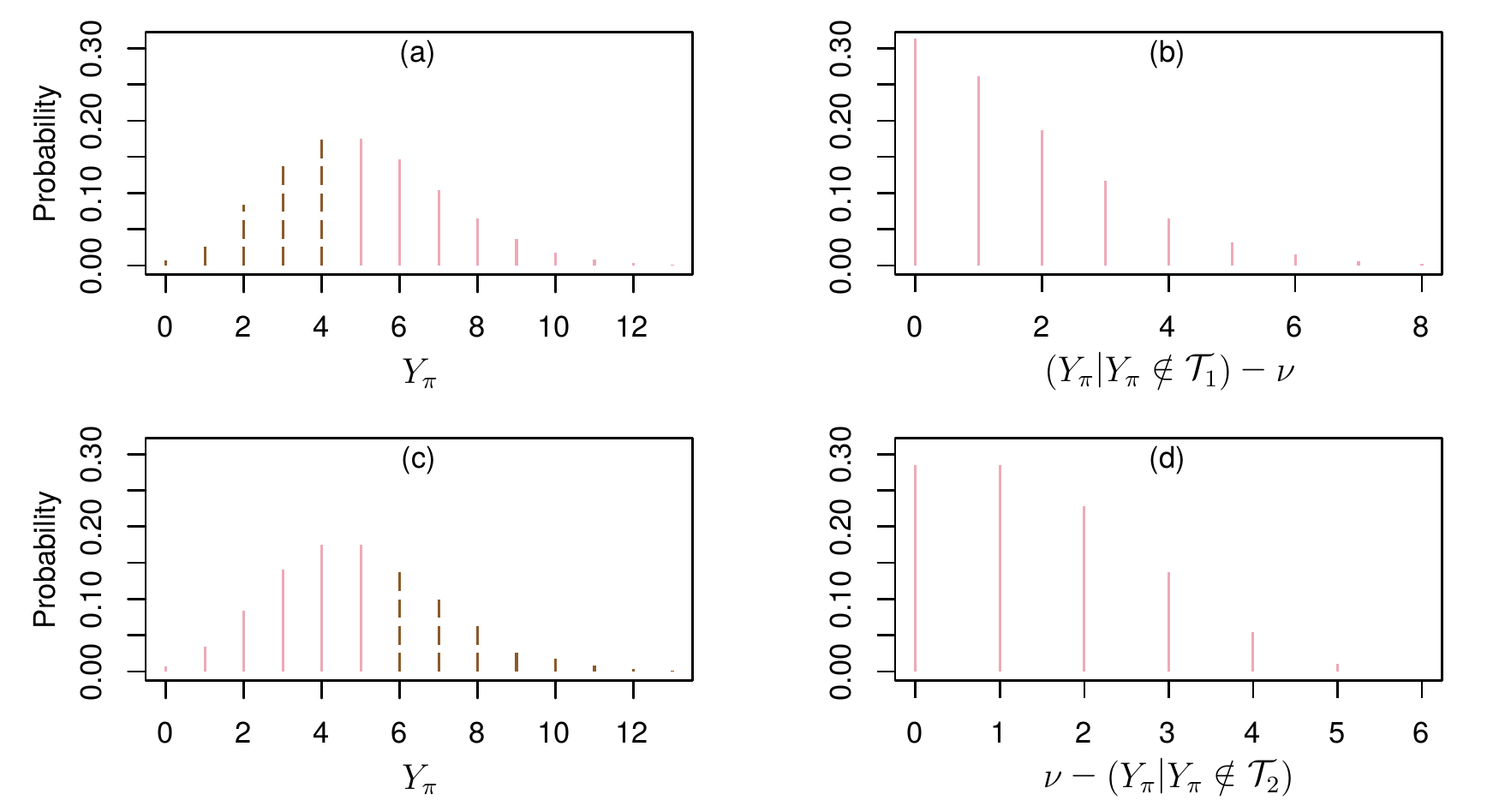}}
\end{center}
\caption{Two generally-truncated Poisson distributions
with amount of shift~$\nu = 5$.
The parent PMF~$f_{\pi}$
is for~$Y_{\pi} \sim \mathrm{Pois}(\lambda_{\pi}=5)$.
(a)~$f_{\pi}$ with $\calT_1=\{0,1,2,3,4\}$ is dashed,
(b)~$Y_* = Y_{\pi} - \nu \sim$ GT--$f_{\pi}$--$\calT_1$
matches the upper tail,
(c)~$f_{\pi}$ with $\calT_2=\{6,7,\ldots\}$ dashed,
(d)~$Y_* = \nu - Y_{\pi} \sim $ GT--$f_{\pi}$--$\calT_2$
matches the lower tail.
Tan dashed lines have been truncated, and the pink
solid lines are used for~$Y_*$.
\label{fig:GTpois4}
}
\end{figure}

% ---------------------------------------------------------
%\newpage  % ,,,,,,,,,,,,,,,,,,,,,,,,,,,,,,,,,,,,,,,,,,,,,,,,
\subsection{Two Measures: Kullback--Leibler Divergence and $\Xi$}
\label{sec:GAITD:KLD}

A natural question is:
how can the total effect of alteration, inflation, truncation
and deflation be measured relative to the parent distribution?
Because its formula allows simplification,
a convenient solution is to compute a divergence
measure such as the Kullback--Leibler divergence (KLD)
\citep{kull:leib:1955}.
Denoting the PMFs by~$f(\cdot)$ and~$f_\pi(\cdot)$,
$D_{\mathrm{KL}}(f \ \|\ f_\pi) = \mbox{}$
\begin{eqnarray}
\nonumber
  &&
  \omega_p 
      \left[
     \log \frac{ \omega_p}{
     {\sum\limits_{u \in \scalA_{p}} f_{\alpha}(u)} } +
  \sum\limits_{a \in \scalA_p} \, A_{\alpha}(a) \;
      \log \frac{ f_{\alpha}(a) }{ f_\pi(a) }
      \right]
      +
  \sum\limits_{s=1}^{|\scalA_{np}|} \omega_s \,
      \log \frac{ \omega_s }{ f_\pi(a) }
        +
 \Delta \, \left( \log  \Delta \right) \; \Pr(y \notin \calS)
 +   \mbox{}  ~~~~
  \\  && \nonumber
  \sum\limits_{i \in \scalI_p}
      \left[ \Delta \, f_\pi(i) + \phi_p \, A_{\iota}(i) \right] \;
         \log  \left\{
         \Delta + \phi_p \, \frac{ A_{\iota}(i) }{ f_\pi(i) }
         \right\}
      +
  \sum\limits_{s=1}^{|\scalI_{np}|}
      \left[ \Delta \, f_\pi(i_s) + \phi_s \right] \;
      \log \left\{ \Delta + \frac{\phi_s }{ f_\pi(i_s) }  \right\}
      + \mbox{}
  \\  &&
  \sum\limits_{d \in \scalD_p}
      \left[ \Delta \, f_\pi(d) - \psi_p \, A_{\delta}(d) \right] \;
         \log \left\{
         \Delta - \psi_p \, \frac{ A_{\delta}(d) }{ f_\pi(d) }
         \right\}
      +
  \sum\limits_{s=1}^{|\scalD_{np}|}
      \left[ \Delta \, f_\pi(d_s) - \psi_s \right] \;
  \log \left\{ \Delta - \frac{ \psi_s }{ f_\pi(d_s) } \right\} ~~~~
  \label{eq:gaitd.KL}
\end{eqnarray}
as $0\cdot \log 0 \equiv 0$ by a limit argument, where
$
\Pr(y \notin \calS) =
1 - \omega_p - \phi_p - \psi_p -
    \sum_{t \in \scalT} f_\pi(t) -
    \sum_{u=1}^{| \scalA_{np} |} \omega_u -
    \sum_{u=1}^{| \scalI_{np} |} \phi_u -
    \sum_{u=1}^{| \scalD_{np} |} \psi_u,
$
and
$A_{\alpha}(y) =
{f_{\alpha}(y)} / {\sum\limits_{u \in \scalA_{p}} f_{\alpha}(u)}$,
$A_{\iota}(y) =
{f_{\iota}(y)} / {\sum\limits_{u \in \scalI_{p}} f_{\iota}(u)}$, etc.

Rather than using the KLD, a  related question is:
what proportion of the data is heaped or seeped?
It is proposed that the approximate measure
$\Xi := \mbox{}$
\begin{eqnarray}
  \sum\limits_{a \in \scalA_{p}}
  \left| \Delta f_{\pi}(a) - 
  \frac{\omega_p \, f_{\alpha}(a)}{
  \sum\limits_{u \in \scalA_p} \; f_{\alpha}(u)}
\right| +
  \sum\limits_{s = 1}^{|\scalA_{np}|}
  \bigg| \Delta f_{\pi}(a_s) - \omega_s \bigg| +
\max \! \left(
\phi_p + \sum\limits_{s = 1}^{|\scalI_{np}|} \phi_s, ~
\psi_p + \sum\limits_{s = 1}^{|\scalD_{np}|} \psi_s \right)  ~~
\label{eq:gaitd.pheapseep} 
\end{eqnarray}
based on~(\ref{eq:dgaitdpmf.combo})--(\ref{eq:dgaitdpmf.combo.mux})
be used which measures the discrepancy between the GAITD PMF
and the scaled parent distribution
on~$\calS \backslash \calT$.
Being opposites, the maximum of total inflation and deflation
is taken.

% ,,,,,,,,,,,,,,,,,,,,,,,,,,,,,,,,,,,,,,,,,,,,,,,,
%\newpage   % ,,,,,,,,,,,,,,,,,,,,,,,,,,,,,,,,,,,,,,,,,,,,,,,,
\subsection{GT--Expansion Method for Underdispersed Data}
\label{sec:GAITD:GTExpansion}

Even armed with the flexibility afforded by general truncation
alone, the following is the first of two useful GAITD special
cases.

Intuitively,
the \textit{Generally-Truncated--Expansion} (GTE) method
combats underdispersion by
a count-preserving transformation that
increases the spread at a greater rate than the mean.
This is achieved by simply multiplying the response by
some integer~$m\ (>1)$ and generally-truncating the values
in between, e.g., doubling~$Y$ and truncating odd values.
By choosing integer~$m$,
the expanded response remains integer-valued.
Writing~$Y \sim (\mu,\ \sigma^2)$ for the first moments,
then~$mY \sim (m\mu,\ m^2\sigma^2)$
has a variance-to-mean ratio (VMR or dispersion index)
of~$m\sigma^2/\mu$.
Since~$\sigma^2/\mu <1$ for underdispersed data,
there exists sufficiently large~$m$
whereby the VMR $\geq 1$.
Ideally the aim is to transform~$Y$ to equidispersion,
or near equidispersion if possible,
and apply the GAITD--Poisson.
If not, then transform to
mild to moderate
overdispersion and use the GAITD--NB.
As the expansion factor~$m$ is not unique, it is
suggested that the smallest value achieving equidispersion
or overdispersion be used.
Further motivation for the GTE method derives from there being
more distributions for handling overdispersion compared to
underdispersion \citep{sell:morr:2017}.

Expressing this more formally, consider an
inverse location--scale transformation
\begin{equation}
  \label{eq:gtmux}
 Y_{**} ~=~ \nu + m Y_{*},  \ \ \ ~
m\in \mathbb{Z}^{+}, \ \ ~
\nu \in \mathbb{Z},
\end{equation}
where usually the multiplier $m>1$ is small
and~$\nu$ nonnegative.
The distribution has its support
shifted to the right by~$\nu$ after being
expanded and separated by~$m$,
hence by generally-truncating the values in between,
underdispersion can be handled.
If~$\nu=0$ then
$Y_{**} \sim$ GT--$f_{\pi}(\btheta_{\pi})$
with $\calT = \mathbb{Z}^{+} \, \backslash \, \{m, 2m, \ldots\}$.
Values for $m$ and~$\nu$ may be known or
estimated.
As an example, using the Poisson as $f_{\pi}$,
if $\nu=0$ then
the sample variance and mean are
$m^2 s_{y_*}^2$ and $m \overline{y_*}$,
hence
yields the moment estimator
\begin{eqnarray}
  \label{eq:m.moment.GT-LS}
  \widehat{m} & = &
 {\overline{y_*}} / {s_{y_*}^2}.
\end{eqnarray}
In practice one would round this and/or choose an
integer $> \widehat{m}$.
The method suggests that if the dispersion index $\in (\frac23, 1)$
then the amount of underdispersion is too slight to require
adjustment to equidispersion.
When fitting the VGLM a~$\log m$ offset is needed
because~$\log \mu_{**} = \log (m \mu_{*})$.
The method is illustrated by handling counts undispersed relative to
the Poisson in
Section~\ref{sec:GAITD:eg.xs.nz.sleep}.

% ,,,,,,,,,,,,,,,,,,,,,,,,,,,,,,,,,,,,,,,,,,,,,,,,
%\newpage   % ,,,,,,,,,,,,,,,,,,,,,,,,,,,,,,,,,,,,,,,,,,,,,,,,
\subsection{GT  for Contiguous Segments}
\label{sec:GAITD:GTcontiguous}

A second useful GAITD special case is based on the ability of
general-truncation to allow selection of any single contiguous
segment of a parent distribution's PMF to be fitted to data.
It is~(\ref{eq:gtmux}) with~$m=\pm 1$ and~$\nu \in \mathbb{Z}$.
For example, Fig.~\ref{fig:GTpois4}(a)--(b)
shows the upper tail of a Poisson$(\lambda_{\pi}=5)$
being used by
having~$\calT_1=\{0,1,\ldots,\nu-1\}$ for~$\nu=5$
so that~$Y_* + \nu \sim $
GT--Pois($\lambda_{\pi};\ \calT_1=\{0,1,\ldots,\nu-1\}$).
That is, the RHS of a Poisson distribution is selected
to be the complete distribution
(after appropriate scaling)
of a right-skewed data set.

In the Fig.~\ref{fig:GTpois4}(c)--(d) example
a reflection is used: $\nu - Y_{*} \sim $
GT--Pois($\lambda_{\pi};\ \calT_2=\{\nu+1, \nu+2, \ldots\}$)
where~$\nu=5$.
That is, the LHS of a Poisson distribution is selected
to be the complete distribution of an upper-truncated data set
that happens to have a rough half-normal shape.

\begin{figure}[pp]
\begin{center}
 {\includegraphics[width=0.90\textwidth]{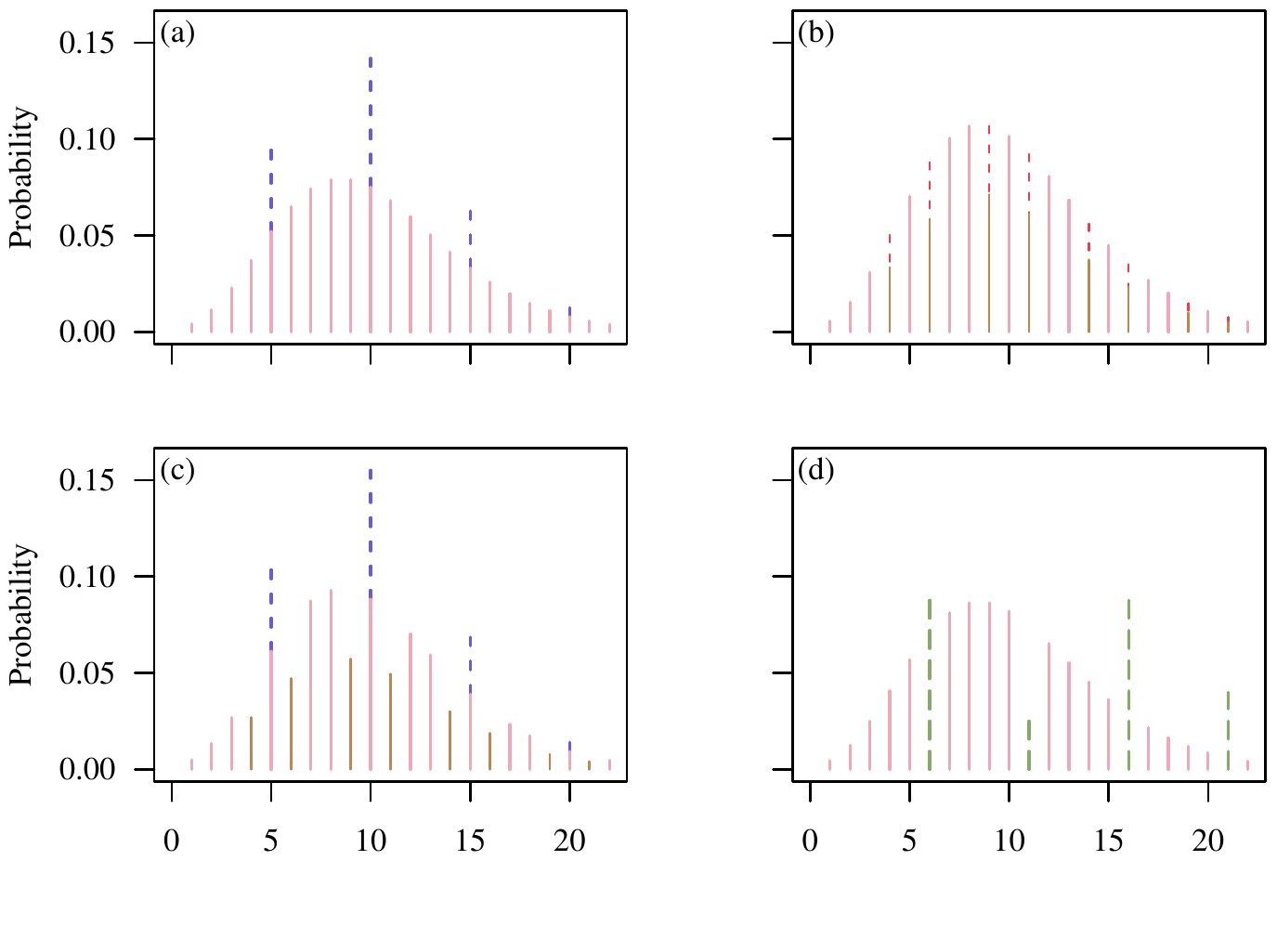}}
\end{center}
\caption{
Heaped and/or seeped data---idealized forms in (a)--(c).
The NBD is the parent.
The special values are
$\calI_p = \{5, 10, 15, 20\}$,
$\calD_{p} = \{4, 6, 9, 11, 14, 16, 19, 21\}$,
$\calT = \{0\}$,
with
$\omega_p = \psi_p = 0.15$,
$\mu_{\pi} =  k_{\pi} =  10$
so that $f_{\pi}=f_{\iota}=f_{\delta}$
are the NB(10, 10) PMF.
(a)~GIT--NB--NB;
(b)~GTD--NB--NB with the dip probabilities shown;
(c)~GITD-NB-NB-NB combines them together;
(d)~GAT-NB-MLM($\bomega_{np}=(0.09, 0.03, 0.09, 0.04)^T$).
The parent distribution comprises the pink solid lines,
the excess probabilities are indigo dashed spikes,
the dip probabilities are reddish dashed lines, and
the altered probabilities are artichoke-colored dashed lines.
\label{fig:Cpoz:dgaitplot.NBD.mix}
}
\end{figure}

% ---------------------------------------------------------
%\newpage  % ,,,,,,,,,,,,,,,,,,,,,,,,,,,,,,,,,,,,,,,,,,,,,,,,
\subsection{Heaped and Seeped Data}
\label{sec:GAITD:heaped.heaping}

Figure~\ref{fig:Cpoz:dgaitplot.NBD.mix}(a)--(c) illustrates
heaping and seeping based on a NB parent,
where all plots have~$\calT=\{0\}$.  Inflation
occurs at multiples of~5 and are shown by dashed indigo lines
appearing as spikes in~(a).  Deflation occurs at the nearest
surrounding values in~(b) where the dip probabilities are the
dashed lines.  The combined plot~(c) is the idealized
heaping-and-seeping scenario because the deficit probabilities are
morphed into excess probabilities.
Somewhat unrelated, plot~(d) illustrates a GAT-NB-MLM where the
dashed artichoke-coloured lines lie arbitrarily below or above the
scaled~$f_{\pi}$.

In contrast to excesses in
Fig.~\ref{fig:Cpoz:dgaitplot.NBD.mix}(a), we describe the deficits
in Fig.~\ref{fig:Cpoz:dgaitplot.NBD.mix}(b) as due to
\textit{seeped} data because some of the nominal parent
probabilities have oozed out, so to speak, due to measurement error.
Strictly speaking, it may be argued that
heaping and seeping form an if-and-only-if relationship:
they appear together or not at all
because spikes must come from dips, and vice versa.
Because there is a conservation of probabilities such as
$\phi_{\lceil i \rceil} = \psi_{\lceil i-1 \rceil} + \psi_{\lceil i+1 \rceil}$
for an inflated value~$i$,
modelling them ideally requires the same variant,
i.e., $\calI_{p}$ and~$\calD_{p}$,
else~$\calI_{np}$ and~$\calD_{np}$.
However, in practice, it would be too strong an assumption
to expect this for all~$i$ in a given data set.
As mentioned in Section~\ref{sec:GAITD:intro.motivation}
the adjacency holds at~$i=30$ years in
Fig.~\ref{fig:xsnz.smokedur.df},
however it does not appear to hold at~$i=25$ or~35.

There is now a sizeable literature on heaped data occurrence.
\cite{craw:weis:such:2015} mention
a wide range of examples taken from
self-reported smoking rates,
duration of breastfeeding,
household total expenditure,
number of drug partners
and
age at menopause.
Not adjusting for heaping in regression can result in biased
estimation
\citep{wang:heit:2008}.
In other situations heaping may lead to
underestimation of within-subject variability
\citep{wang:shiff:griff:heit:2012}.
Despite
\citet[p.572]{craw:weis:such:2015} stating that
``inference for heaped data is an important statistical problem''
we opine that
most techniques proposed have been unduly complex
and indeed,
among the c.18,500 \RR{} packages currently on CRAN, there
appears only one directly focussed on heaped data
(\pkg{Kernelheaping}).
In contrast, we believe that GAITD regression is more accessible
and flexible than competing methods.

% ,,,,,,,,,,,,,,,,,,,,,,,,,,,,,,,,,,,,,,,,,,,,,,,,
%\newpage   % ,,,,,,,,,,,,,,,,,,,,,,,,,,,,,,,,,,,,,,,,,,,,,,,,
\subsection{Special Cases}
\label{sec:GAITD:specialcases}

We take the opportunity to comment on some special cases.
The GA--$f_{\pi}$--MLM and GA--$f_{\pi}$--$f_{\alpha}$ are a
(discrete) spliced or composite distribution since the mixture
components form the
partition~$\calA_p \cup (\calR \backslash \calA_p)$ of~\calR.  In
contrast, GI--$f_{\pi}$--MLM and GI--$f_{\pi}$--$f_{\iota}$ are
not strictly spliced distributions because their component
supports are~$\calI_{p}$ and~\calR{} with the former nested within
the latter
(\cite{su:fan:levi:tan:trip:2013} propose a GI-like model).
In a comprehensive review of mixed Poisson
distributions \citep{karl:xeka:2005} the idea of a spliced or
partially spliced distribution is not mentioned among the many
members cited.  To the best of our knowledge GAITD regression
comprising interlacing discrete spliced and partially spliced
distributions is novel.

One can easily constrain $\btheta_{\pi}=\btheta_{\alpha}$
and $\btheta_{\pi}=\btheta_{\iota}$
in both GAT--$f_{\pi}$--$f_{\alpha}$ and
GIT--$f_{\pi}$--$f_{\iota}$ models
using the constraint matrices of~(\ref{eq:constraints.VGLM}).
As a simple example,
for GAT--Pois($\lambda_{\pi}$)--Pois-($\lambda_{\alpha}$)
the constraint $\lambda_{\pi} = \lambda_{\alpha}$ can be enforced by
\[
\bH_{k} =
\left(
\begin{array}{ccc}
1 & 0 & 1 \\
0 & 1 & 0
\end{array}
\right)^T,
\qquad k=1,\ldots,d,
\]
in~(\ref{eq:constraints.VGLM}) because
$\boldeta=
(\log\,\lambda_{\pi},\ \logit\,\omega_p,\ \log\,\lambda_{\alpha})^T$
if~$| \calA_p | > 1$.
The software easily allows this.
A likelihood ratio test is valid for formally testing for equal
rate parameters.

% =========================================================
%\newpage   % ,,,,,,,,,,,,,,,,,,,,,,,,,,,,,,,,,,,,,,,,,,,,,,,,
\section{Joint Under-- and Over-dispersion Analysis}
\label{sec:GAITD:overdispersion}

The \textit{separate} effects of 0-alteration, 0-inflation,
and 0-truncation
on overdispersion in Poisson regression
has been considered by a number of authors,
e.g.,
\cite{rido:hind:deme:2001},
\cite{tang:lu:chen:etal:2015},
\cite{sell:raim:2016}
and
\cite{hasl:parn:hind:mora:2022},
however GAITD distributions allow a more general and \textit{joint}
investigation of this phenomenon from the four operators.
Indeed,
there is an almost plethora of possibilities.
It is found more convenient to use the
\textit{variance-to-mean difference} (VMD)
$\Var \, Y - \mathrm{E} \, Y$
rather than the VMR.
Determining overdispersion
is exacerbated
by having at least two possible definitions,
with both having their merits:
\begin{eqnarray}
\label{eq:overd.defn.VMD.star}
\mathrm{VMD}_{*}: \quad &
\Var \, Y_{*} - \mathrm{E} \, Y_{*} ~>~ 0,
\\
\mathrm{VMD}_{\pi}: \quad &
\label{eq:overd.defn.VMD.pi}
\Var \, Y_{*} - \mathrm{E} \, Y_{\pi} ~>~ 0.
\end{eqnarray}
The former
transfers the
VMD comparison
onto the new distribution while
the latter compares the variance of the modified distribution
to parent mean.
They
primarily are only applicable to the Poisson,
hence more generally define the
\textit{variance-to-variance difference} as
\begin{eqnarray}
\label{eq:overd.defn.VVD}
\mathrm{VVD}_{}: \quad &
\Var \, Y_{*} - \Var \, Y_{\pi} ~>~ 0
\end{eqnarray}
to replace~(\ref{eq:overd.defn.VMD.pi}).
This compares the variances of the modified and parent distributions
and can be applied to distributions such as the binomial
(the VMD${}_{\pi}$ and VVD coincide for the Poisson.)

Alternatively to the above
perhaps it is better to quantify
overdispersion
more symmetrically based on both pairs of the first two moments:
\begin{eqnarray}
\label{eq:overd.defn.DVMDD}
  \mathrm{DVMD} &=&
  \Var \, Y_{*} - \mathrm{E} \, Y_{*} -
  \left( \Var \, Y_{\pi} - \mathrm{E} \, Y_{\pi} \right)
  ~>~  0,
  \\
  \label{eq:overd.defn.DIR}
  \mathrm{DIR} &=&
  \frac{
  \Var \, Y_{*} / \mathrm{E} \, Y_{*} }{
  \Var \, Y_{\pi} / \mathrm{E} \, Y_{\pi}
  }
  ~>~  1,
\end{eqnarray}
where the acronyms are self-explanatory.
For the Poisson
the VMD${}_{*}$ and DVMD coincide,
as does the DIR and VMD${}_{\pi}$ because the denominator is unity.

Possibly the most tractable method for studying overdispersion in
Poisson regression while allowing for the joint effects of the
four operators is the following result.

\medskip

\noindent
\textbf{Theorem}
\quad
\textit{(i)~Under the VMD${}_{*}$~(\ref{eq:overd.defn.VMD.star})},
\textit{overdispersion relative to the Poisson distribution
will occur for the
GAITD--$f_{\pi}$--$f_{\alpha}$--MLM--$f_{\iota}$--MLM--$f_{\delta}$--MLM
combo if}
\begin{eqnarray}
\nonumber
\lefteqn{
\frac{ \omega_{p} \,
\sum\limits_{a \in \scalA_{p}} \, a(a-1) \, f_{\alpha}(a)
}{
\sum\limits_{a \in \scalA_{p}} \, f_{\alpha}(a)
}
+
\frac{ \phi_{p} \,
\sum\limits_{i \in \scalI_{p}} \, i(i-1) \, f_{\iota}(i)
}{
\sum\limits_{i \in \scalI_{p}} \, f_{\iota}(i)
}
-
\frac{ \psi_{p} \,
\sum\limits_{d \in \scalD_{p}} \, d(d-1) \, f_{\delta}(d)
}{
\sum\limits_{d \in \scalD_{p}} \, f_{\delta}(d)
}
+
\mbox{}
} \\
\nonumber
\lefteqn{
\sum_{a \in \scalA_{np}} \, a(a-1) \, \omega_{\lceil a \rceil} +
\sum_{i \in \scalI_{np}} \, i(i-1) \, \phi_{\lceil i \rceil} -
\sum_{d \in \scalD_{np}} \, d(d-1) \, \psi_{\lceil d \rceil} +
\mbox{}
} \\
&&
\Delta \; \cdot
\left\{
\mathrm{E}[Y_{\pi}(Y_{\pi}-1)]
- \sum\limits_{a \in \{\scalA_p,\; \scalA_{np}\}} \, a(a-1) \, f_{\pi}(a)
- \sum\limits_{t \in \scalT} \, t(t-1) \, f_{\pi}(t)
\right\}
~>~
\mu_*^2,
\label{eq:gaitcombo.pois.overd.VMD.star}
\end{eqnarray}
\textit{where $\mu_*$ is given by~(\ref{eq:gait.combo.EYk})
with $k=1$};
\textit{(ii)~Likewise,
  under the VMD${}_{\pi}$~(\ref{eq:overd.defn.VMD.pi})}
\textit{if}
\begin{eqnarray}
\nonumber
\lefteqn{
\frac{ \omega_{p} \,
\sum\limits_{a \in \scalA_{p}} \, (a-\mu_{*})^2 \, f_{\alpha}(a)
}{
\sum\limits_{a \in \scalA_{p}} \, f_{\alpha}(a)
}
+
\frac{ \phi_{p} \,
\sum\limits_{i \in \scalI_{p}} \, (i-\mu_{*})^2 \, f_{\iota}(i)
}{
\sum\limits_{i \in \scalI_{p}} \, f_{\iota}(i)
}
-
\frac{ \psi_{p} \,
\sum\limits_{d \in \scalD_{p}} \, (d-\mu_{*})^2 \, f_{\delta}(d)
}{
\sum\limits_{d \in \scalD_{p}} \, f_{\delta}(d)
}
+
\mbox{}
} \\
\nonumber
\lefteqn{
\sum_{a \in \scalA_{np}} \, (a-\mu_{*})^2  \, \omega_{\lceil a \rceil} +
\sum_{i \in \scalI_{np}} \, (i-\mu_{*})^2 \, \phi_{\lceil i \rceil} -
\sum_{d \in \scalD_{np}} \, (d-\mu_{*})^2 \, \psi_{\lceil d \rceil} +
\mbox{}
} \\
&&
\Delta \; \cdot
\left\{
\sigma_{\pi}^2 + (\mu_{\pi} - \mu_{*} )^2
- \sum\limits_{a \in \{\scalA_p,\; \scalA_{np}\}}\, (a-\mu_{*})^2 \,f_{\pi}(a)
- \sum\limits_{t \in \scalT} \, (t-\mu_{*})^2  \, f_{\pi}(t)
\right\}
~>~
\mu_{\pi};  ~~~
\label{eq:gaitcombo.pois.overd.VMD.pi}
\end{eqnarray}
\textit{(iii)~Likewise, under the VVD~(\ref{eq:overd.defn.VVD}),
if~(\ref{eq:gaitcombo.pois.overd.VMD.pi})
but with~$\sigma_{\pi}^2 = \Var \, Y_{\pi}$ replacing~$\mu_{\pi}$
on the RHS}.

\noindent \textbf{Proof} \quad
Overdispersion for VMD${}_{*}$
follows from $\mathrm{E}[Y_{*}(Y_{*}-1)] > \{\mathrm{E}(Y_{*})\}^2$.
The other cases are straightforward and use
$\mathrm{E} \left[ \left( Y_{\pi} - \mu_{*} \right)^2 \right] =
\Var( Y_{\pi} ) + (\mu_{\pi} - \mu_{*} )^2$.
\hfill{$\Box$}

\bigskip

\noindent
\textbf{Corollary}
\quad
\textit{For nondegenerate 0-altered
  count distributions overdispersion with
  respect to~VMD${}_{*}$ occurs when}
\begin{eqnarray}
  \label{eq:GAITD.VMD.pi.za}
  \frac{1 - \omega}{1 - f_{\pi}(0)}
  \left[ \,
  \sigma_{\pi}^2 - \mu_{\pi} +
  \frac{\mu_{\pi}^2 \, \{ \omega - f_{\pi}(0) \}}{1 - f_{\pi}(0)}
  \right] ~>~ 0.
\end{eqnarray}
\textit{So for the Poisson distribution overdispersion occurs
when~$\Pr(Y_{\pi} = 0) < \omega$,
i.e., 0 is heaped,
and underdispersion when~0 is seeped.}

\begin{figure}[tt]
\begin{center}
 {\includegraphics[width=0.7\textwidth]{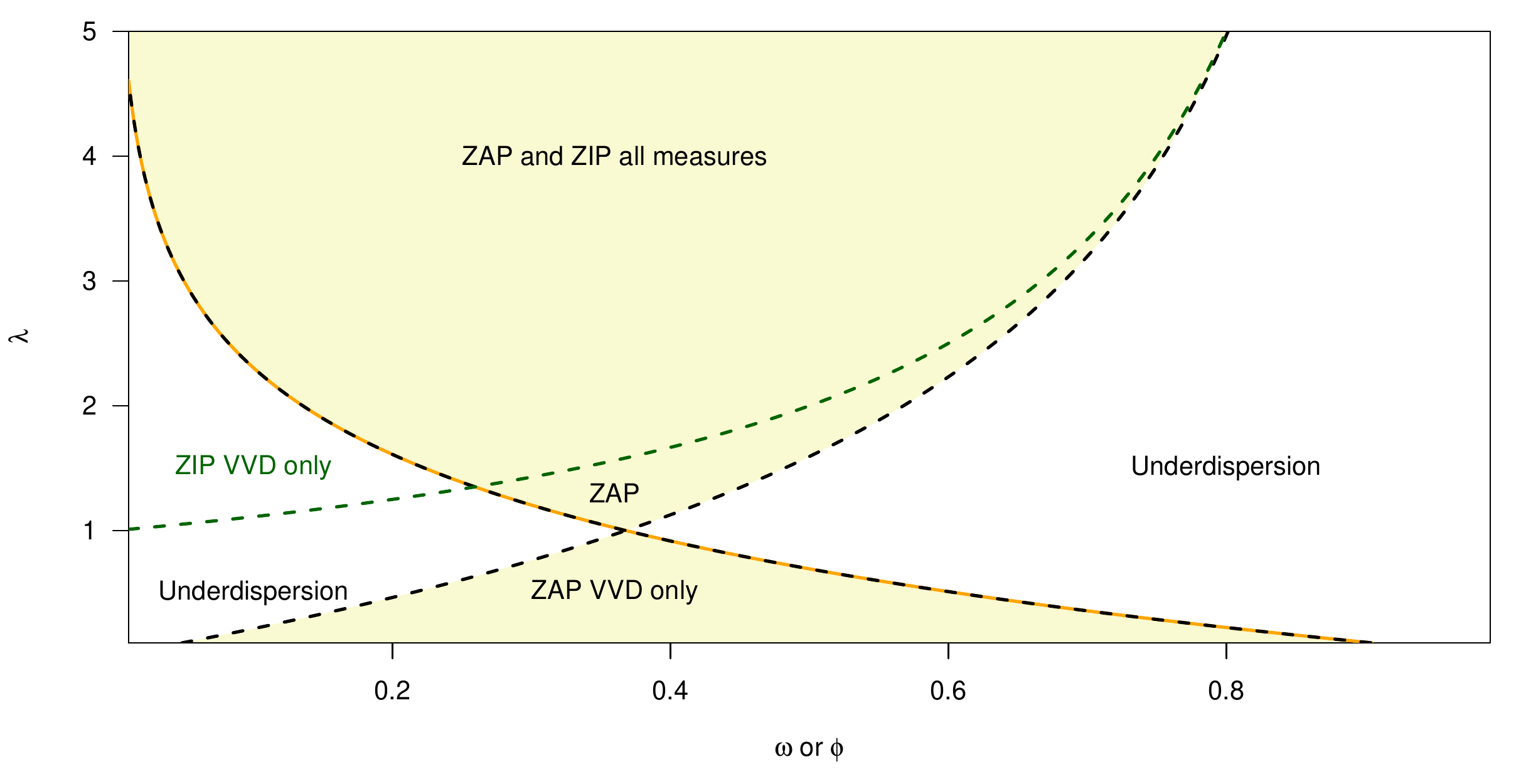}}
\end{center}
\caption{
Regions in the parameter space where
ZA- and ZI-Poisson($\lambda_{\pi}$) regressions
are overdispersed relative to the Poisson, according to
definitions~(\ref{eq:overd.defn.VMD.star})--(\ref{eq:overd.defn.DIR}).
Approximately, the parameter space
is~$\lambda_{\pi} \in (0.1, 5)$
and~$\omega$ and~$\phi \in (0, 1)$.
The ZAP VVD is shaded.
\label{fig:GAITP:gaitoverdPois1}
}
\end{figure}

\medskip

According to the various definitions, it is interesting to see
the conditions for which overdispersion occurs for
two common 1-parameter count distributions with~0
as the special value.
Table~\ref{tab:GAITD.overdispersion} is a summary of
the Poisson and binomial cases.
Several points emerge, e.g.,
\begin{itemize}\itemsep0pt

\item[(i)]
the different definitions do yield different conditions
for overdispersion.
Zero-truncation does not result in overdispersion by any measure.

\item[(ii)]
The VVD tends to produce the most complicated conditions
because overdispersion occurs in two disjoint regions
of the parameter space.
However, overdispersion does not occur
for the ZAP when~$\lambda = 1$ and~$\omega = e^{-1}$,
nor does it occur
for the ZAB when~$p = 1 / (N+1)$ and~$\omega = (1 - p)^{N}$;
in these cases equidispersion occurs.

\item[(iii)]
Zero-inflation tends to result in overdispersion more than alteration.

\item[(iv)]
Zero-alteration usually occurs when~$\omega$ is too large,
which contrasts with
zero-inflation which usually occurs when~$\phi$ is too small.

\end{itemize}
Fig.~\ref{fig:GAITP:gaitoverdPois1} displays
partitions the parameter space for the Poisson parent
with~$0.1 < \lambda_{\pi} < 5$ and~$0 < \omega / \phi < 1$.
Overdispersion occurs in regions above the curves, and in the
case of the ZAP-VVD it also occurs in the region wedged in
at the central bottom.
Excluding zero-truncation,
overdispersion tends to increase with increasing~$\lambda_{\pi}$
because the distribution shifts
away from the origin to leave the point mass at~0
for creating extra variation.

\renewcommand{\arraystretch}{1.1}
\begin{table}[tt]
\caption{
Summary of conditions for overdispersion under
definitions~(\ref{eq:overd.defn.VMD.star})--(\ref{eq:overd.defn.DIR}):
the 0-altered, 0-inflated and 0-truncated cases are considered
for the Poisson and binomial parent distributions where
$0 < \lambda_{} < \infty$,
$0 < p_{} < 1$,
$0 < \omega < 1$ and
$0 < \phi   < 1$
are assumed.
Here,
$q = 1-p$,
$\phi = \Pr(Y=0)$ structurally and
$\omega = \Pr(Y=0)$; and
$\dagger$ indicates the endpoints of the interval may need
switching to preserve the order.
\label{tab:GAITD.overdispersion}
}
\centering
\begin{tabular}{|lcccc|}
  \hline
$f_{}$ &
VMD${}_{*}$ &
VVD &
DVMD &
DIR \\
  \hline
ZAP($\lambda_{},\ \omega$) &
$e^{-\lambda_{}} < \omega$
&
$
\omega \in
\displaystyle{
\left[ e^{-\lambda_{}},\ 1 - \frac{1 - e^{-\lambda_{}}}{\lambda} \right]
} \, \dagger ~~
$
      &
$e^{-\lambda_{}} < \omega$
      &
$e^{-\lambda_{}} < \omega$
\\
ZAB($N,\ p_{},\ \omega$) ~ &
$q^N +
\displaystyle{
\frac{1-q^N}{N} < \omega
}$ ~
& ~
$
\omega \in
\displaystyle{
\left[ q^N,\ 1 - \frac{q (1 - q^N)}{N (1 - q)} \right]
} \, \dagger ~
$
&
~ $
\displaystyle{
q^N
< \omega}$ ~~
&
$q^N < \omega$
\\
ZIP($\lambda_{},\ \phi$) &
Always &
$\phi < \displaystyle{ 1 - \frac{1}{\lambda_{}} }$
&
Always
&
Always
\\
ZIB($N,\ p_{},\ \phi$) ~~~~&
Always &
$\phi < 1 - \displaystyle{ \frac{1-p}{N p} }$
       &
Always
       &
Always
\\
ZTP($\lambda_{}$) &
Never  &
Never
&
Never
&
Never
\\
ZTB($N,\ p_{}$) ~~~~&
Never &
Never &
Never &
Never
\\
\hline
\end{tabular}
\end{table}
\renewcommand{\arraystretch}{1.0}

% =========================================================
%\newpage  % ,,,,,,,,,,,,,,,,,,,,,,,,,,,,,,,,,,,,,,,,,,,,,,,,
\section{Maximum Likelihood Estimation}
\label{sec:GAITD:estimation}

The technical details and derivative systems for
maximizing~$\ell$ by Fisher scoring/IRLS
are given in the Supplementary Materials;
here we give commentary on a few overarching details.

The full parameter space for GAITD regression is described by
the last equations of~(\ref{eq:dgaitdpmf.combo})
and~(\ref{eq:dgaitdpmf.combo.constraints1})
coupled with~(\ref{eq:dgaitdpmf.combo.mux}).
Since~$\Delta > 0$ implies
\[
  0 ~<~
  \omega_p + \phi_p 
   + \sum\limits_{u=1}^{| \scalA_{np} |}  \omega_u
   + \sum\limits_{u=1}^{| \scalI_{np} |}  \phi_u
   ~<~
   1 + \psi_p
   + \sum\limits_{u=1}^{| \scalD_{np} |}  \psi_u,
\]
it is seen that the parameter space boundary is not fixed,
hence this dependency breaches a standard regularity condition.
A consequence is
that~$\mathrm{E}[\partial \ell / \partial \psi_p] = 2\neq  0$
for example.
Instead, we estimate the probabilities in~(\ref{eq:gaitd.etas})
by an ordinary MLM.  Operating in a reduced parameter
space may occasionally create inconvenience
because the probability of the baseline reference group,
$\Pr(y \notin \calS)$, may become perilously close to~0, e.g.,
when there is much nonparametric inflation or deflation so that
the sum of the~$\phi_s$ and~$\psi_s$ is large.
In contrast, it would have been ideal if deflation could release
probability back into the model
that could be used for inflation, however this is not possible.
Using the MLM to estimate the model means that the
parameter space used is~$1/[1 + 2\phi_p + 2\sum \psi_v]$
the size of the `proper' parameter space.
We call~$\Pr(y \notin \calS)$ the \textit{nonspecial baseline
  probability} (NBP) or \textit{reserve} probability.  Because it
must be positive it pays to be specify the~$\psi_s$ economically so
that one does not exhaust the reserve probability unnecessarily.

There will be the occasional dataset exhibiting
much inflation and deflation.
The following strategies can help economize the
probability-consumption problem caused by MLM estimation.
\begin{enumerate}\itemsep0pt

\item
  Parametric inflation/deflation is more economical;
  can some~$\phi_s$/$\psi_s$ be represented by~$\phi_p$/$\psi_p$?
  For example, $\psi_s \equiv \psi_p \, f_{\delta}(y)$.

\item
  If~$\psi_s > \frac12 f_{\pi}(y)$ then use~$\omega_s$ instead,
i.e., alter~$y$ rather than deflating it.

\end{enumerate}

% =========================================================
%\clearpage  % ,,,,,,,,,,,,,,,,,,,,,,,,,,,,,,,,,,,,,,,,,,,,,,,,
%\newpage  % ,,,,,,,,,,,,,,,,,,,,,,,,,,,,,,,,,,,,,,,,,,,,,,,,
\section{Examples}
\label{sec:GAITD:egs}

Data for both examples are subsets from
a large ($n=10,529$) New Zealand cross-sectional study
(and an approximate random sample of the country's working
population then)
collected in the mid-1990s \citep{macm:etal:1995}.
Code for reproducing the analyses
are included in the Supplementary Materials.

\begin{figure}[tt]
\begin{center}
{\includegraphics[width=0.75\textwidth]{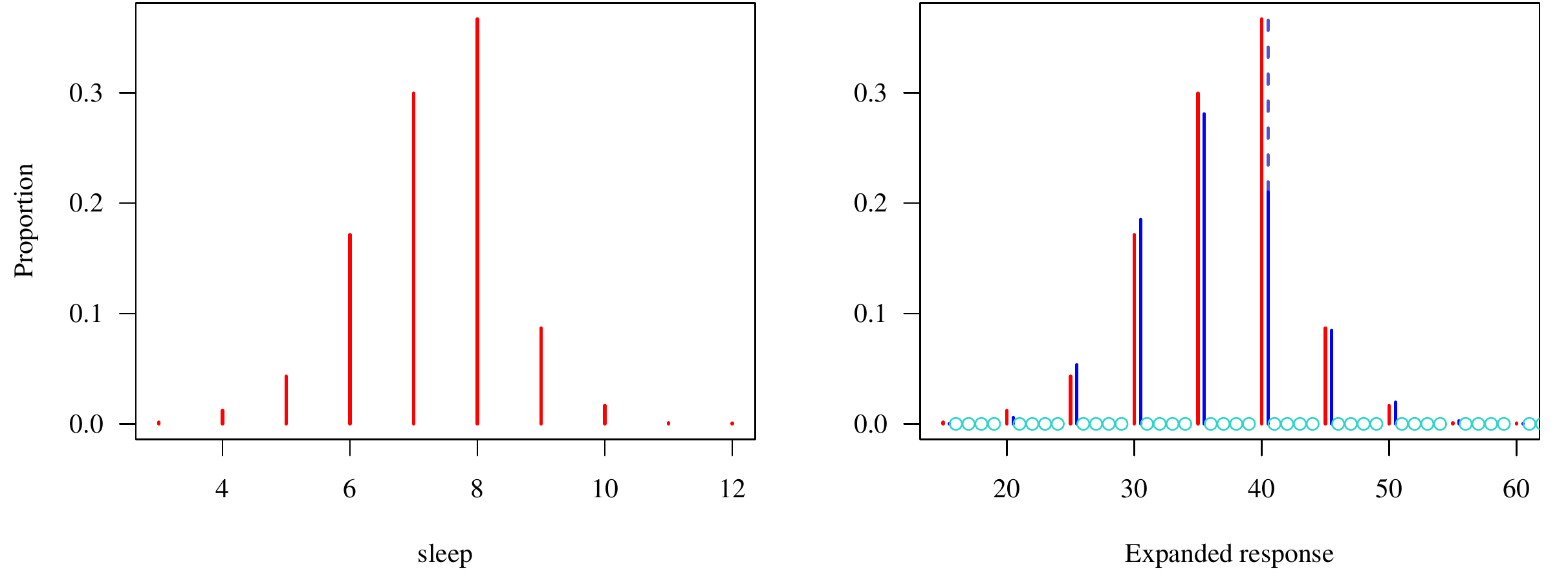}}
\end{center}
\caption{
(a)~Spikeplot of \texttt{sleep} in \texttt{xs.nz}
from \pkg{VGAMdata};
(b)~GT-Expansion method with GIT-Poisson fit is
overlaid adjacently in blue---the
truncated values are turquoise hollow points.
\label{fig:spikeplot-xsnz-sleep}
}
\end{figure}

% ---------------------------------------------------------
%\newpage   % ,,,,,,,,,,,,,,,,,,,,,,,,,,,,,,,,,,,,,,,,,,,,,,,,
\subsection{Sleep Duration}
\label{sec:GAITD:eg.xs.nz.sleep}

To simply illustrate the GT-Expansion method
we consider the self-reported sleep duration response to
the question
``How many hours do you usually sleep each night?''
All recorded answers were integer-valued,
and after removing the missing values and outliers ($2.4\%$)
there were~$n=10,264$ individuals;
specifically,
$y \in \{3, \ldots, 12\}$ so we
chose~$\calT = \{0, 1, 2, 13, 14, \ldots\}$
to account for the removals and physical limits.

The data (Table~\ref{tab:xs.nz.sleep}) comprises of~10 values
and might be considered only marginally heaped
because there are no obvious spikes even
though the data are clearly tainted by measurement error.
A routine Poisson regression is not amenable because:
(i)~there is strong underdispersion
(the sample mean and variance are~$7.3$ and~$1.3$);
(ii)~left-skew is apparent
(Fig.~\ref{fig:spikeplot-xsnz-sleep}(a))
while the Poisson tends to be right-skewed.
We combat these by applying the GTE method and
assigning~$\calI_{np}=\{8\}$.
The 8-inflation might be justified because of
the common belief that about~8 hours sleep is
recommended for most adults
\citep[e.g.,][]{hirs:etal:2015}.

\renewcommand{\arraystretch}{0.6}
\begin{table}[tt]
\caption{
Usual sleep duration in a New Zealand cross-sectional data.
\label{tab:xs.nz.sleep}
}
\centering
\begin{tabular}{|lrrrrrrrrrr|}
\hline
Hours &
   3 &  4 &  5 &  6  & 7 &  8 &  9 & 10 & 11 & 12 \\
Frequency  &
  16& 125 &443&1760&3076&3766 &891 &170 & 10  & 7 \\
\hline
\end{tabular}
\end{table}
\renewcommand{\arraystretch}{1.0}

A simple search over all multipliers
yields~$m=5$ for maximizing the GIT likelihood.
This compares to the moment estimator~(\ref{eq:m.moment.GT-LS})
which is~$5.67$.
Fig.~\ref{fig:spikeplot-xsnz-sleep}(b)
overlays the GIT model on the data and
a good fit to the observed proportions is seen.
The model indicates that the amount of inflation
is~$\widehat{\phi}_{\lceil 8 \rceil} \approx 0.157$---almost~1/6
of the entire data set.
The Kullback-Leibler divergence
from the model to an ordinary Poisson is~$4.57$.
The overall GIT mean, (\ref{eq:gait.combo.EYk}),
is estimated by~$7.297$ hours
whereas an approximate~95\% confidence interval
for~$\mu_{\pi}$
is~$[7.139, 7.194]$ hours.
The former is higher because the 8-inflation draws the overall mean
towards it.

%==============================================================
% \clearpage  % ,,,,,,,,,,,,,,,,,,,,,,,,,,,,,,,,,,,,,,,,,,,,,,,,,
% \newpage  % ,,,,,,,,,,,,,,,,,,,,,,,,,,,,,,,,,,,,,,,,,,,,,,,,,
\subsection{Smoking Duration}
\label{sec:GAITD:eg.xs.nz.smokeyears}

This example
concerns how long current smokers or ex-smokers reported smoking
in years.
A small fraction of the data
(about~$0.2$\%)
had values~0.1, 0.2, 0.3 and~0.5
which were rounded,
as well as~$2.5$\%
that were missing values and were deleted.
The positive integer-valued smoking duration data set
has~$n=5492$ with
almost half having never smoked
(about~$48$\%).
Of those who do,
Fig.~\ref{fig:xsnz.smokedur.df}
shows a heavy-tailed distribution of smoking duration
with one or two layers of heaping that is
unimodal with mean between~10 and~20 years.
The most pronounced heaped values
include~$\calI_{p}= \{ 5, 10, 20, 30, 40, 50, 60 \}$,
as well as~12, 25, 35.
A careful examination also suggests
that~$\calD_{p}= \{9, 11, 13, 19, 21, 29, 31 \}$ are seeped.
Furthermore, we chose
$\calA_{p}  = \{ 2, 15, 25, 35, 45 \}$,
$\calI_{np} = \{ 1, 8, 12, 18 \}$,
and a NB parent to handle overdispersion.
We relaxed the assumptions that the altered and inflated
distributions are equal to the parent.

\begin{figure}[tt]
\begin{center}
 {\includegraphics[width=0.9\textwidth]{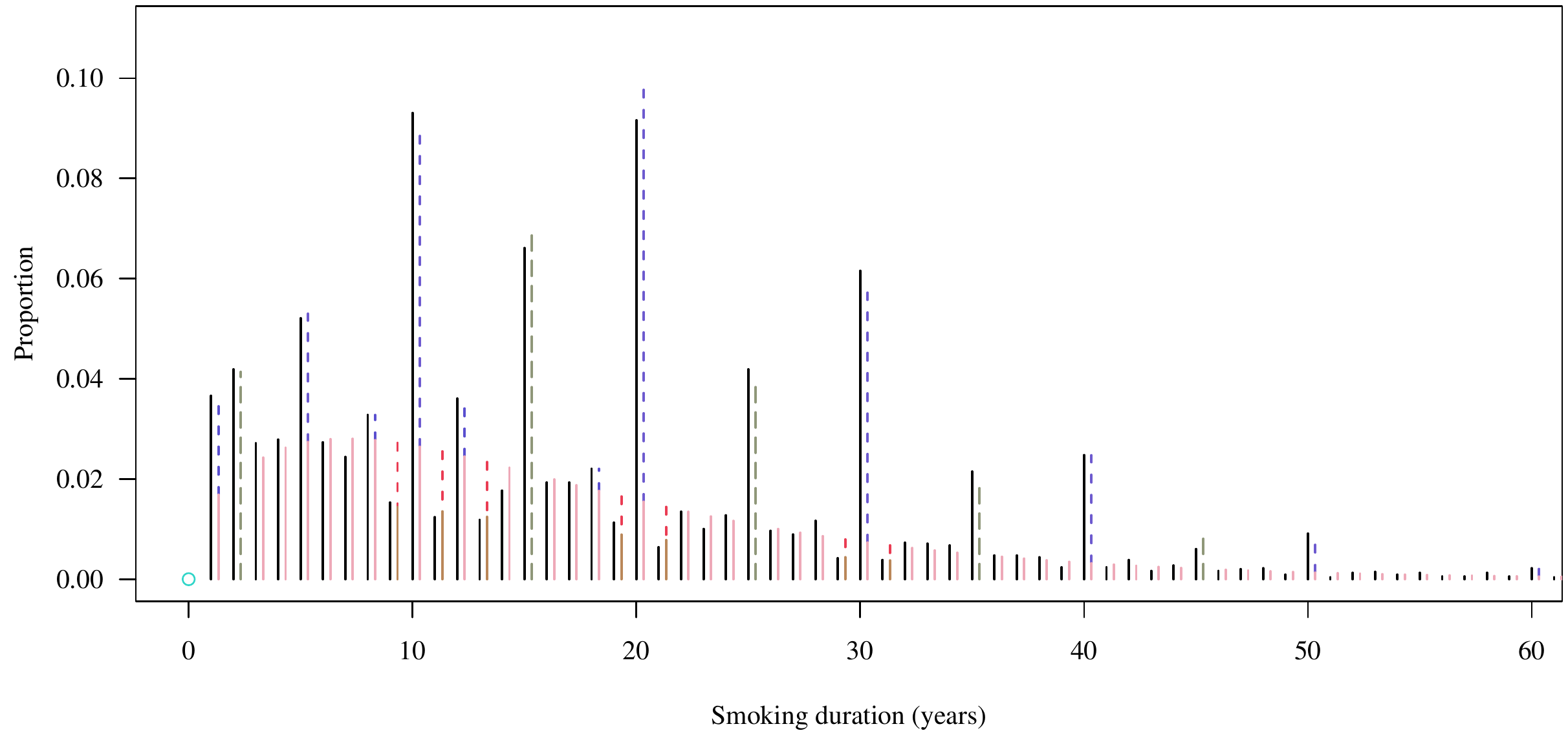}}
\end{center}
\caption{
  How \texttt{smokeyears} in Fig.~\ref{fig:xsnz.smokedur.df}
  and the GAITD regression compare.
\label{fig:spikeplot4-smokedf}
}
\end{figure}

We first fitted an intercept-only GAITD regression.
Because the nonsmokers were such a large group it was necessary
to truncate~0 to conserve the baseline probability:
$\calT_{  } = \{ 0 \}$.
The assumptions that the altered and inflated
distributions are equal to the parent
was relaxed.
Fig.~\ref{fig:spikeplot4-smokedf}
shows a very good correspondence between the model and data.
To conserve the baseline reserve probability, $\calI_p$
was used to model the layer of largest spikes
while~$\calA_p$ for the inner layer.

Our model showed almost every regression coefficient being very
significant (p-value $< 0.001$)
and~$\Xi \approx 41$\% of the data was heaped.
A rootogram of the fit indicated the response residuals having no
systematic lack-of-fit, hence we concluded that the model fitted
well to these data.

Next, we added sex and ethnicity to our model.
Details of the final model are  placed in
the Supplementary Materials.
Transcribing the coefficients into linear predictors,
the model~(\ref{eq:gaitd.etas}) has~$\widehat{\boldeta}= 
(\widehat{\eta}_1,\ldots,\widehat{\eta}_{15})^T =  \mbox{}$
$
  \bigg(
  \log \, \widehat{\mu}_{\pi} =
  2.736
  + \mathrm{I}(\mathtt{Male}) \, 0.124
  - \mathrm{I}(\mathtt{Maori}) \, 0.262
  - \mathrm{I}(\mathtt{Polynesian}) \, 0.238
  - \mathrm{I}(\mathtt{Other}) \, 0.217, 
  \ \log \, \widehat{k}_{\pi}  =  0.631, ~
g(\widehat{\omega}_p)  =  -0.976,~
  \log \, \widehat{\mu}_{\alpha}  =
  2.839
  + \mathrm{I}(\mathtt{Male}) \, 0.134
  - \mathrm{I}(\mathtt{Maori}) \, 0.185
  - \mathrm{I}(\mathtt{Polynesian}) \, 0.315
  - \mathrm{I}(\mathtt{Other}) \, 0.450,
  \ \log \, \widehat{k}_{\alpha}  =  0.806,~
g(\widehat{\phi}_p)  =  -0.622,
  \ \log \, \widehat{\mu}_{\iota}  =
  3.118
  + \mathrm{I}(\mathtt{Male}) \, 0.041
  - \mathrm{I}(\mathtt{Maori}) \, 0.137
  - \mathrm{I}(\mathtt{Polynesian}) \, 0.189
  - \mathrm{I}(\mathtt{Other}) \, 0.395,
  \ \log \, \widehat{k}_{\iota}  =  1.476,~
  g(\widehat{\psi}_p)  =  -2.110,
  \ \log \, \widehat{\mu}_{\delta}  =  \widehat{\eta}_1,
  \ \log \, \widehat{k}_{\delta}  =  \widehat{\eta}_2,
\ g(\widehat{\phi}_{\lceil 1  \rceil})  =  -3.112,~
g(\widehat{\phi}_{\lceil  8 \rceil})  =  -4.508,~
g(\widehat{\phi}_{\lceil 12 \rceil})  =  -3.716,~
g(\widehat{\phi}_{\lceil 18 \rceil})  =  -4.660
\bigg)^T$
where~$g$ is the multinomial logit link.
In particular,
the final model suggests that:
(i)~Europeans smoke longer than the other three ethnicities---and
there appears little difference between the three;
(ii)~Males smoke longer in general,
however there does not seem to be a difference between
males and females in the inflated values (spikes).

% =========================================================
% \newpage  % ,,,,,,,,,,,,,,,,,,,,,,,,,,,,,,,,,,,,,,,,,,,,,,,,
\section{Discussion}
\label{sec:GAITD:discussion}

This paper extends models commonly known by the abbreviations
ZIP, ZAP, ZTP, ZDP, ZINB, ZAB, ZTB, \ldots, into a
maximal class of models having
four operators with parametric and nonparametric variants
operating concurrently.
The resultant combo model provides much needed unification.
However, even with the limited additional flexibility afforded
by having~\{0\} as the only special value, such models
have been found easily misused, and
guidelines have evolved for fitting them.
For example,
some authors of
software have highlighted common
mistakes made by practitioners,
e.g.,
the \pkg{mgcv}~1.8-38 \texttt{ziP} helpfile
\citep{wood:2017}
urges users about checking
and cautions about identifiability problems and convergence warnings.
These guidelines transfer across to GAITD regression with
even greater force,
and since GAITD regression is so flexible, it is likely
that overfitting will be a common error among
novice users.

Another area where practitioners need to exhibit more care
is hypothesis testing,
e.g., $H_0: \phi = 0$ for the ZIP.
Then the usual regularity conditions do not hold
for ordinary likelihood inference
at the boundaries
and special measures need to be adopted, e.g.,
\cite{mora:1971},
\cite{self:lian:1987}.
Many practitioners adopt the \cite{vuon:1989} test,
e.g.,
\citet[pp.574--6,863]{greene:2012},
however its recommendation is not universal \citep{wils:2015}.
With GAITD regression the issue
is aggravated by having multiple boundaries
from the mixing probabilities of GI models.

The choice of the~$i_s$ warrants comment.  As generally-inflated
models can easily be overused, selection of the elements of
$\calI_{p}$ and $\calI_{np}$ should ideally be justified prior to
the data being examined, as well as empirical observation of
unequivocal excess values.  For example, in analyses of~20
multivariate data sets of count data, \cite{wart:2005} concluded
that it was rarely necessary to fit 0-inflated models to data sets
with high frequencies of~0s when the estimated NB (empirically,
the best fitting distribution) mean parameter was low.  To quote
\pkg{mgcv}: `Zero inflated models are often over-used.
Having lots of zeroes in the data does not in itself imply zero
inflation. Having too many zeroes \textit{given the model mean}
may imply zero inflation.'  For GAITD regression any support value
considered inflated should be justified in the context of the
entire distribution and parameter values.

A possible consequence of the GT-expansion method is a reduction
of the
need for developing parametric distributions to handle
underdispersion, for example, the Conway--Maxwell--Poisson
distribution
\citep[e.g.,][]{sell:borl:shmu:2012,huan:2017.cmp}
has seen a revival but is beset by computational difficulties.

We mention two avenues for future work in closing.
Firstly, this work very naturally extends to continuous
distributions where spikes are augmented by slabs because
alteration, inflation and deflation can
operate on subintervals.
Some preliminary work has already commenced in this area.
Secondly,
several layers of altered/inflated/deflated values are conceivable,
which would entail~$\calA_{1p}, \calA_{2p}, \ldots$,
and~$\calI_{1p}, \calI_{2p}, \ldots$,
and~$\calD_{1p}, \calD_{2p}, \ldots$
replacing~$\calA_{p}$, $\calI_{p}$ and~$\calD_{p}$.

\section*{Supplementary Materials}

The computational and software details placed in the
supplementary materials are currently unavailable online.
The software implementation is in the \pkg{VGAM} package
(version~1.1-6 onwards) on CRAN;
please consult the online help for details.

\section*{Acknowledgements}

We thank Luca Frigau and Alan Huang for useful feedback,
Paul Murrell and Simon Urbanek for help with the figures,
Theodora Ge Jin for support,
Rolf Turner for help with the writing,
and delegates of the Multivariate Count Analysis workshop held
at Besan\c{c}on, France, in July 2018 for helpful
feedback---especially
Dimitris Karlis for bringing his work to our attention.
CM was supported by a 2018 University of Auckland Northern
Hemisphere Summer Research Scholarship while a student
at Zhejiang University.

\bibliographystyle{plainnat}
\bibliography{allgaitd.bib}

%-------------------------------------------
\bigskip
\bigskip
\bigskip

\vskip .65cm
\noindent
Thomas~Yee,
\vskip 2pt
\noindent
Department of Statistics, University of Auckland, New Zealand.
\vskip 2pt
\noindent
E-mail: t.yee@auckland.ac.nz

\bigskip

\noindent
Chenchen Ma
\vskip 2pt
\noindent
    School of Mathematical Sciences and
    Center for Statistical Science,
    Peking University, China.
\vskip 2pt
\noindent
E-mail: ChenchenMa@pku.edu.cn

\end{document}